\documentclass[a4paper,fleqn,usenatbib]{mnras}

\usepackage{mathptmx}

\usepackage[T1]{fontenc}
\usepackage{ae,aecompl}

\usepackage{graphicx}	
\usepackage{amsmath}	
\usepackage{amssymb}	
\usepackage{hyperref}
\usepackage{caption}
\usepackage{multirow}
\usepackage{pdflscape}
\usepackage{makecell}
\usepackage{txfonts}

\newcommand{\HST}{{\it HST}}

\newcommand{\OII}{{\rm [OII]}}

\newcommand{\hal}{{\rm H$\alpha$}}
\newcommand{\hb}{{\rm H$\beta$}}
\newcommand{\logM}{{\rm log$_{10}(M_{*}$/M$_{\odot}$)}}

\defcitealias{2000ApJ...533..682C}{C00}
\defcitealias{2019ApJS..243...22B}{B19}
\defcitealias{2004AJ....127.2002K}{K04}
\defcitealias{1998ARA&A..36..189K}{K98}

\title[Differential attenuation in star-forming galaxies at 0.3\,$\lesssim$\,$z$\,$\lesssim$\,1.5]{Differential attenuation in star-forming galaxies at 0.3\,$\lesssim$\,$z$\,$\lesssim$\,1.5 in the SHARDS/CANDELS field}
\author[L. Rodr\'{\i}guez-Mu\~noz et al.]{L. Rodr\'{\i}guez-Mu\~noz$^{1}$\thanks{E-mail: lucia.rdguez.munoz@gmail.com},
G. Rodighiero$^{1}$,
P. G. P\'erez-Gonz\'alez$^{2}$,
M. Talia$^{3,4}$,
\newauthor I. Baronchelli$^{1,5}$,
L. Morselli$^{1}$,
A. Renzini$^{5}$,
A. Puglisi$^{6}$,
A. Grazian$^{5}$,
A. Zanella$^{5}$,
\newauthor C. Mancini$^{7,1}$,
A. Feltre$^{4}$,
M. Romano$^{1,5}$,
A. Vidal Garc\'{\i}a$^{8}$,
A. Franceschini$^{1}$,
\newauthor B. Alcalde Pampliega$^{9}$,
P. Cassata$^{1}$,
L. Costantin$^{2}$,
H. Dom\'{\i}nguez S\'anchez$^{10}$,
\newauthor N. Espino-Briones$^{11}$,
E. Iani$^{12}$,
A. Koekemoer$^{13}$,
A. Lumbreras-Calle$^{14}$,
\newauthor J.M. Rodr\'{i}guez-Espinosa$^{15,16}$
\\
$^{1}$Dipartimento di Fisica e Astronomia ``G. Galilei'', Universit\`a degli Studi di Padova, Vicolo dell'Osservatorio 3, I-35122, Padova, Italy\\
$^{2}$Centro de Astrobiolog\'{\i}a, Instituto Nacional de T\'ecnica Aeroespacial, Carretera de Ajalvir km\,4, Torrej\'on de Ardoz, E-28850, Madrid, Spain\\
$^{3}$University of Bologna - Department of Physics and Astronomy, Via Gobetti 93/2, I-40129, Bologna, Italy\\
$^{4}$INAF - Osservatorio di Astrofisica e Scienza dello Spazio, Via Gobetti 93/3, I-40129, Bologna, Italy\\
$^{5}$INAF - Osservatorio astronomico di Padova, Vicolo Osservatorio 5, I-35122, Padova, Italy\\
$^{6}$Center for Extragalactic Astronomy, Durham University, South Road, Durham DH1 3LE, United Kingdom\\
$^{7}$INAF - Istituto di Astrofisica Spaziale e Fisica Cosmica Milano, via Bassini 15, I-20133, Milano, Italy\\
$^{8}$LPENS, École Normale Supérieure, Université PSL, CNRS, Sorbonne Université, Université Paris-Diderot, 75005, Paris, France\\
$^{9}$European Southern Observatory, Alonso de C\'ordova 3107, Casilla 19001, Vitacura, Santiago, Chile\\
$^{10}$Institute of Space Sciences (ICE, CSIC), Campus UAB, Carrer de Magrans, E-08193, Barcelona, Spain\\
$^{11}$Departamento de F\'{\i}sica de la Tierra y Astrof\'{\i}sica, Facultad de CC F\'{\i}sicas, Universidad Complutense de Madrid, E-28040, Madrid, Spain\\
$^{12}$Kapteyn Astronomical Institute, University of Groningen, 9700AV, Groningen, The Netherlands\\
$^{13}$Space Telescope Science Institute, 3700 San Martin Dr., MD 21218, Baltimore, USA\\
$^{14}$Centro de Estudios de Física del Cosmos de Aragón, Plaza San Juan 1, R-44001, Teruel, Spain\\
$^{15}$Instituto de Astrofísica de Canarias, E-38205, La Laguna, Spain\\
$^{16}$Depto. de Astrofísica, Universidad de La Laguna, E-38206, La Laguna, Spain\\
}

\date{Accepted XXX. Received YYY; in original form ZZZ}

\pubyear{2021}

\begin{document}
\label{firstpage}
\pagerange{\pageref{firstpage}--\pageref{lastpage}}
\maketitle

\begin{abstract}
We use a sample of 706 galaxies, selected as {\OII$\lambda$3727} (\OII) emitters in the Survey for High-$z$ Absorption Red and Dead Sources {(SHARDS)} on the {CANDELS/}GOODS-N field, to study the differential attenuation of the nebular emission with respect to the stellar continuum. The sample includes only galaxies with a counterpart in the infrared and \logM\,$>$\,9, over the redshift interval 0.3\,$\lesssim$\,$z$\,$\lesssim$\,1.5. Our methodology consists in the comparison of the star formation rates inferred from  \OII\,and \hal\, emission lines with a robust quantification of the total star-forming activity ({\it SFR}$_{\mathrm{TOT}}$) that is independently estimated based on both infrared  and ultraviolet (UV) luminosities. We obtain $f$$=$$E(B-V)_{\mathrm{stellar}}$/$E(B-V)_{\mathrm{nebular}}$\,$=$\,0.69$^{0.71}_{0.69}$ and {0.55$^{0.56}_{0.53}$} for \OII\, and \hal, respectively. {Our resulting $f$-factors display a significant positive correlation with the UV attenuation and shallower or not-significant trends with the stellar mass, the {\it SFR}$_{\mathrm{TOT}}$, the distance to the main sequence, and the redshift}.  Finally, our results favour an average nebular attenuation curve similar in shape to the typical dust curve of local starbursts.
\end{abstract}

\begin{keywords}
galaxies: star formation --
galaxies: ISM --
galaxies: evolution -- 
galaxies: high-redshift --
ISM: dust, extinction
\end{keywords}



\section{Introduction}
Dust attenuation has a strong impact on the shape of the spectral energy distribution (SED) of galaxies. Dust absorbs and scatters photons at short wavelengths and thermally emits the absorbed energy in the infrared (IR; $\lambda\sim$ 1-1000\,$\mu$m; e.g., \citealt{2007ApJ...657..810D}). In this way, dust causes a major uncertainty in the derivation of galaxy physical properties (e.g., stellar mass, star formation rate) either through SED-modeling techniques or direct use of luminosity measurements. This is particularly problematic at high redshift ($z$$\gtrsim$2), where the detailed characterization of galaxy populations frequently relies on the rest-frame ultraviolet (UV) and optical wavelength regimes (\citealt{2014ARA&A..52..415M}).

Extinction and attenuation curves describe the impact of dust on the emission of galaxies as a function of the wavelength. The former describe how the light is absorbed or scattered out of the line-of-sight (e.g., \citealt{1989ApJ...345..245C}, \citealt{1999PASP..111...63F}), whereas the latter account also for the scattering of light into the line-of-sight and for a non-uniform distribution of dust column densities (e.g., \citealt{2000ApJ...533..682C}, hereafter C00). Attenuation curves are shaped by the complex interplay between the properties of dust grains and the spatial distribution of dust and stars within galaxies (e.g., \citealt{2001PASP..113.1449C}, \citealt{2020ARA&A..58..529S}). 

Stellar continuum attenuation curves have been derived using theoretical modeling (\citealt{2000ApJ...539..718C}) and empirical approaches using local galaxy samples (e.g., \citetalias{2000ApJ...533..682C}) and high-redshift galaxy samples. Several of these works have identified an important variability of the attenuation curves as well as dependencies with galaxy physical properties (e.g., \citealt{2018ApJ...869...70N}, \citealt{2018ApJ...859...11S}, \citealt{2019MNRAS.488.2301T}, \citealt{2019ApJS..243...22B}, \citealt{2021MNRAS.502.3210L}, \citealt{2020ApJ...899..117S}, \citealt{2020ApJ...903..146B}, \citealt{2021ApJ...909..213K}). In literature, the most frequently used prescription to model the average impact of dust in the stellar emission of galaxies out to high redshift is the attenuation curve by \citetalias{2000ApJ...533..682C}. {However, recent  evidence  shows that  steeper dust  curves  may   be more representative of high-redshift galaxy samples ($z$$\gtrsim$2, e.g.,  \citealt{2018ApJ...853...56R}, 
\citealt{2019ApJ...871..128T}, 
\citealt{2020MNRAS.491.4724F}).}

The dust impact on nebular emission arising in ionized gas in star-forming regions of galaxies appears to be somewhat different to that affecting the stellar continuum. In particular, a number of studies of both local (\citealt{1988ApJ...334..665F}, \citealt{1997AJ....113..162C}, \citetalias{2000ApJ...533..682C}, \citealt{2011MNRAS.417.1760W},  \citealt{2013ApJ...771...62K}) and high-redshift star-forming galaxies (e.g., \citealt{2009ApJ...706.1364F},  \citealt{2010ApJ...718..112Y}, 
\citealt{2011ApJ...738..106W}, 
\citealt{2013ApJ...777L...8K}, 
\citealt{2013ApJ...771...62K},  \citealt{2013ApJ...779..135W},  \citealt{2014ApJ...788...86P},  \citealt{2015ApJ...806..259R},  \citealt{2015A&A...582A..80T},  \citealt{2016ApJ...820...96D},  \citealt{2018A&A...619A.135B}, \citealt{2020ApJ...899..117S}) have found that emission lines (EL) are subject to a higher attenuation than the stellar continuum. This effect is generally referred to as {\it differential attenuation}, and it is expected considering a two component dust-star spatial distribution model, in which the population of young stars together with the nebular emission they trigger are embedded in molecular clouds with higher dust covering fractions than that of the diffuse interstellar medium (ISM) that the non-ionizing stellar continuum is mainly affected by (e.g., \citealt{1994ApJ...429..582C}, \citealt{2000ApJ...539..718C}). This dust spatial distribution model could imply not only an excess of {reddening} for the ELs but also a different shape of the nebular {dust} curve. Frequently, either the Galactic extinction curve (\citealt{1989ApJ...345..245C}) or the attenuation curve of local starbursts by \citetalias{2000ApJ...533..682C} are adopted for the nebular line emission up to high redshift (\citealt{2020ApJ...902..123R}).  

In spite of the great efforts, no general consensus has been reached on the nature and magnitude of the differential attenuation. While some authors find large differences between the {reddening} of nebular and stellar emission (e.g., \citetalias{2000ApJ...533..682C}, \citealt{2009ApJ...706.1364F};  \citealt{2013ApJ...777L...8K}, \citealt{2013ApJ...771...62K}, \citealt{2014ApJ...788...86P},
\citealt{2015A&A...582A..80T}), 
others find ELs suffering from reddening at a similar rate as the continuum (e.g., \citealt{2013ApJ...777L...8K}, \citealt{2006ApJ...647..128E},  \citealt{2007ApJ...670..156D}, \citealt{2010ApJ...712.1070R}, \citealt{2012ApJ...744..154R}, \citealt{2015ApJ...804..149S}, \citealt{2016A&A...586A..83P}). {Sample selection differences might be at the origin of these discrepancies, which suggests that the relation between the stellar and nebular reddening likely depends on galaxy properties.}

Over the last decades, a large amount of new data from multi-wavelength imaging and spectroscopic surveys has revealed the properties of large samples of star-forming galaxies (SFG). Among other spectral features, ELs are powerful tools to identify SFGs out to high redshifts, and characterize their properties in terms of star formation (SF) and physics of their ISM (e.g., metallicity and excitation; \citealt{2019ARA&A..57..511K}). {However, the majority of the brightest and most thoroughly studied ELs belong to a wavelength range in which the impact of dust is significant} (e.g., \hal\,and \OII $\lambda\lambda$3726,3729 doublet, hereafter \OII). Still, ELs are key for the ongoing and future wide and deep spectroscopic and photometric surveys such as the J-PLUS photometric survey (\citealt{2019A&A...622A.176C}), the Extended Baryon Oscillation Spectroscopic Survey (eBOSS, \citealt{2016AJ....151...44D}), the Dark Energy Spectroscopic Instrument (DESI, the \citealt{2016MNRAS.460.1270D}), Euclid (\citealt{2011arXiv1110.3193L}), the Wide Field Infrared Survey Telescope (WFIRST; \citealt{2012arXiv1210.7809D}) and,
Subaru Prime Focus Spectrograph Survey (\citealt{2014PASJ...66R...1T}). These projects will unveil a large number of SFGs out to high redshift by detecting mainly their rest-frame UV and optical ELs. Thus, understanding how dust shapes nebular emission is crucial for the correct characterization of emission line galaxy (ELGs) samples throughout cosmic times, and the interpretation of the results in the global context of galaxy evolution.

The objective of this work is to explore the imprints of dust in the nebular emission of SFGs, and to that end, we use the outstanding wealth of data available on the GOODS-N field of the Cosmic Assembly Near-infrared Deep Extragalactic Legacy Survey (CANDELS; \citealt{2011ApJS..197...35G}, \citealt{2011ApJS..197...36K}). In particular, we use the ultra-deep spectrophotometric Survey for High-z Absorption Red and Dead
Sources (SHARDS; \citealt{2013ApJ...762...46P}) to select a sample of SFGs throughout the redshift range 0.3-1.5 by their \OII-emission. 
This EL has proven to be a useful SFR indicator in absence of hydrogen Balmer lines, in particular H$\alpha$ at $z$$>$0.4 in optical data. Despite the impact of the ISM properties (i.e., metallicity and ionization parameter) in the relation between the ionizing UV luminosity emitted by young stars and the luminosities of forbidden lines, the excitation of \OII\, can be exploited as a SFR indicator (e.g., \citealt{1989AJ.....97..700G}; \citealt{1998ARA&A..36..189K}, \citealt{2002MNRAS.332..283R}, \citealt{2003ASPC..297..191A}, \citealt{2004AJ....127.2002K}, hereafter K04, \citealt{2015A&A...582A..80T}). The multi-wavelength coverage of CANDELS/GOODS-N field includes several bands probing the UV, optical,  {near-infrared (NIR), mid-infrared (MIR), and  far-infrared (FIR)}, as well as ground based and {\it HST}-grism spectroscopy. These data enable the accurate characterization of the samples of SHARDS \OII-emitters (e.g., \citealt{2015ApJ...812..155C}). By combining the SF activity traced by the  \OII\, and \hal\,with those probed by the UV continuum and IR, we derive the average differential {reddening} suffered by \OII\, and \hal\,as a function of redshift and the physical properties of galaxies, giving at the same time insights into the shape of the nebular attenuation curve.

This publication is organized as follows. Section~\ref{sec:data} and~\ref{sec:oiiemitters} present the data and methodology used to identify ELGs in SHARDS.  Section~\ref{sec:flux} reports how ELs (\OII\, and \hal) are measured using the SHARDS spectrophotometric data. In Section~\ref{sec:physprop}, we present a brief characterization of the final sample of \OII-emitters. Section~\ref{sec:ffactor} presents the results regarding the {differential reddening} of \OII~and \hal\,ELs. In Section~\ref{sec:ffactor_dep}, we explore the dependencies of the observed {differential reddening} on physical properties of galaxies and redshift. In Section~\ref{sec:discuss}, we discuss our results, we explore the shape of the nebular attenuation curve, and review the caveats to take into account when studying the differential attenuation. Finally, a summary of our main findings and conclusions can be found in Section~\ref{sec:conclusions}.

Throughout this work we assume a flat $\Lambda$CDM cosmology with $H_0$$=$$70$\,kms$^{-1}$Mpc$^{-1}$, $\Omega_{m}$$=$$0.3$, and $\Omega_{\Lambda}$$=$$0.7$. Stellar masses and {\it SFR} are quoted for a \citet{2003PASP..115..763C} initial mass function (IMF; stellar masses from 0.1 to 100~M$_{\odot}$), and magnitudes are given in the AB photometric system (\citealt{1974ApJS...27...21O}).

\section{Data}\label{sec:data}
This paper makes use of the catalog on the GOODS-N field (\citealt{2004ApJ...600L..93G}) published by \citet[][hereafter B19]{2019ApJS..243...22B}. This catalog provides UV-to-FIR integrated photometry of the 35445 sources detected in the 171~arcmin$^{2}$ WFC3/F160W map of CANDELS. The limiting magnitudes (at a 5$\sigma$ significance) range between 27.8-28.7 mag (within a 0.77" radius aperture) over the  wide, intermediate, and deep regions of the map.
The catalog includes also the physical properties of all the galaxies in CANDELS-GOODS-N field. These properties are derived by fitting the observed SEDs (UV-to-NIR and {MIR-to-FIR} separately) with galaxy emission templates using different SED-fitting codes. The models fitted to the UV-to-NIR SEDs are built with \cite{2003MNRAS.344.1000B} stellar population synthesis models with a \cite{2003PASP..115..763C} IMF, exponentially declining star formation histories (SFH) with a minimum e-folding time of log$_{10}$\,($\tau$/yr)\,$=$\,8.5, a minimum age of 40~Myr, a solar metallicity, an attenuation between 0\,$<$\,A$_V$\,$<$\,4~mag, and the \citetalias{2000ApJ...533..682C} attenuation law. {Several studies favor steeper dust curves for high-redshift galaxies, particularly at $z$$\gtrsim$2 (e.g., \citealt{2006ApJ...644..792R,2018ApJ...853...56R}, 
\citealt{2012ApJ...758L..31L},   \citealt{2013ApJ...772..136O},  \citealt{2016ApJ...831..176B}, 
\citealt{2019ApJ...871..128T}, 
\citealt{2017MNRAS.472..483F,2020MNRAS.491.4724F}). However, \citetalias{2000ApJ...533..682C} seems to successfully model the average impact of dust in large samples of galaxies out to intermediate-to-high redshift (e.g., \citealt{2011ApJ...738..106W}). For instance,  Figure~4 in \citet{2011ApJS..193...13B} displays the residuals between observed (UV and optical) fluxes and synthetic photometry derived from best-fitting templates for a subsample of spectroscopically confirmed SFGs in the Extended Groth Strip field. No systematic deviations appear, except for the wavelength range affected by the 2175\AA\, bump, which plays no role in our results.} Furthermore, the models include emission lines. It is worth noting that \citet{2010MNRAS.407..830M} have shown that an exponentially declining SFH is a poor approximation for the SFH of high redshift galaxies, leading to unrealistically young ages. The derived stellar masses are also affected, by as much as a factor of ~2. However, this mismatch has no major impact on our results. The fitting of the IR regime of the SEDs is performed for galaxies which are detected at a significance level larger than 5$\sigma$ in the \textit{Spitzer}/MIPS 24\,$\mu$m band and at least one of the \textit{Herschel}/PACS and/or SPIRE maps. \citetalias{2019ApJS..243...22B} use the dust emission models published by \citet{2001ApJ...556..562C}, \citet{2002ApJ...576..159D}, \citet{2009ApJ...692..556R}, and \citet{2007ApJ...657..810D}. In the following sections, we further describe the content of the \citetalias{2019ApJS..243...22B} catalogs which are relevant for our work.

\subsection{Broad- and medium-band photometry}
The \citetalias{2019ApJS..243...22B} catalog includes broadband photometry in the UV (U band from KPNO and LBC), optical ({\it HST}/ACS F435W, F606W, F775W, F814W, and F850LP), and IR ({\it HST}/WFC3 F105W, F125W, F140W, and F160W; {\it Subaru}/MOIRCS Ks; CFHT/Megacam K; and {\it Spitzer}/IRAC 3.6, 4.5, 5.8, and 8.0 $\mu$m, {\it Spitzer}/MIPS 24 $\mu$m, {\it Herschel}/PACS 100 and 160 $\mu$m, SPIRE 250, 350 and 500 $\mu$m) bands. {Details on the methodology followed by \citetalias{2019ApJS..243...22B} to measure the photometry on {\it Herschel} bands minimizing the effects of confusion can be found in their Appendix~D.1.1. Briefly, source catalogues and photometry are obtained through a PSF fitting technique relying on {\it Spitzer}/IRAC and MIPS priors. In order to match the FIR fluxes with the {\it HST}/WFC3 F160W source catalogue, \citetalias{2019ApJS..243...22B} apply the methodology described by \citet{2019MNRAS.485..586R}. In practice, they identify the most likely shorter wavelength counterpart to the FIR detections using the information of the whole NIR-to-FIR wavelength range (see \citetalias{2019ApJS..243...22B} Appendix~D.1.2).} 

Furthermore, the catalog includes optical spectrophotometric data from SHARDS (\citealt{2013ApJ...762...46P}). SHARDS is an ESO/GTC Large Program that covered the GOODS-N field  with ultra-deep (220 hours) GTC/OSIRIS (Optical System for Imaging and low-Intermediate-Resolution Integrated Spectroscopy) imaging through 25 medium-band optical filters (see \citetalias{2019ApJS..243...22B} for a description of the dataset). These bands cover a continuous wavelength range from 5000 to 9500\,\AA\,giving the spectral information equivalent to a R$\sim$50 spectrum. The width of the filters ranges from 13.8 to 33.3\,nm. The depth of the imaging reaches 26.5 mag at the 4$\sigma$ level for every single filter, and the seeing remains always below 1\arcsec.  SHARDS used 2 OSIRIS (field of view, FoV, 7.8'$\times$7.8') pointings to cover an area similar to that targeted by CANDELS on GOODS-N. 

\subsection{Spectroscopic data}
\citetalias{2019ApJS..243...22B}'s catalog includes also information gathered from numerous spectroscopic surveys, mainly in the optical and near-IR. A total of $\sim$5000 unique redshifts among which $\sim$3000 are assigned a highly reliable quality flag. Among the different campaigns on the CANDELS/GOODS-N field, we highlight the 3D-HST survey (\citealt{2012ApJS..200...13B}), which provides {\it HST}/WFC3 IR spectroscopic observations with the G102 and G141 grisms (R$\sim$210 and 130, respectively). In this work, we make use of the H$\alpha$ ELs measurements performed by \citet{2016ApJS..225...27M} on 3D-HST spectra. {We only consider H$\alpha$ fluxes with SNR$>$3.}

\subsection{Redshifts}\label{sec:data_z}
The spectroscopic redshifts (spec-$z$) are collected from numerous optical and NIR ground- and space-based surveys (see \citetalias{2019ApJS..243...22B} Section~2.4.1, and references therein).
On the other hand, the photometric redshifts (photo-$z$) given by \citetalias{2019ApJS..243...22B} are computed differently for each source depending on the availability of \HST/WFC3 grism data. When grism data is not available photo-$z$ are obtained using a slightly modified version of the \texttt{EAZY} code (\citealt{2008ApJ...686.1503B}) adapted to take into account the spatial variation in the effective wavelength of the SHARDS filters depending on the galaxy position in the SHARDS mosaics. A modified version of the SED-fitting code developed by the 3D-\HST\,survey and discussed in \citet{2012ApJS..200...13B} and \citet{2016ApJS..225...27M} is used otherwise (see \citetalias{2019ApJS..243...22B} Section 5.1.3 for details). The quality assessment of the photometric redshifts derived using these two methodologies gives fractions of outliers ($\Delta z$/(1+$z$)$>$0.15), 3.3\% and 2.7\%, respectively. On the other hand, the normalized median absolute deviation (\citealt{1983ured.book.....H}) of the difference between the photo-$z$ and the spec-$z$, which is a variable frequently used for the quantification of the scatter in the photo-$z$ $vs$ spec-$z$ plane (e.g., \citealt{2009ApJ...690.1236I}, \citealt{2017MNRAS.470...95M}, \citealt{2019MNRAS.485..586R}), presents values 0.0028 and 0.0023, respectively. This means that the average uncertainty of the photometric redshifts used in this work is $\lesssim$0.003$\times$(1+$z$),
i.e., $\lesssim$\,0.3\%. 

\subsection{Stellar mass}\label{seq:MASS}
The stellar mass ($M_{*}$) of each galaxy is estimated from the average scale factor required to match the template monochromatic luminosities to the observed UV-to-NIR fluxes, weighted with the photometric errors. The random uncertainty of the $M_{*}$ is derived from the dispersion in the mass-luminosity ratios in the different bands. The average expected uncertainty taking into account variations in metallicity, SFH, or IMF is within 0.3~dex (\citealt{2008ApJ...675..234P}).
The stellar mass completeness level of our CANDELS parent catalog at the {highest redshift of interest for our study} is log$_{10}$($M_{*}/$M$_{\odot}$)$\sim$9 (\citetalias{2019ApJS..243...22B}; see also \citealt{2015A&A...575A..96G}).

\subsection{Star formation rate from UV and IR luminosities}\label{seq:SFRCONT}
In order to derive the {star formation rate ({\it SFR})} traced by the UV continuum, \citetalias{2019ApJS..243...22B} use the recipe by \citet[][hereafter K98]{1998ARA&A..36..189K} converted into a \citealt{2003PASP..115..763C} IMF:
\begin{equation}
SFR_{\mathrm{UV,\,K98}}/M_{\odot}yr^{-1} =8.8 \times 10^{-29} L_{\nu, \mathrm{UV}}/\mathrm{erg\,s}^{-1}\mathrm{Hz}^{-1} \label{eq:SFRuv1600}\\
\end{equation}
We note that this expression can be used with monochromatic luminosities within the wavelength range between 1250-2800\AA\, where the stellar UV spectrum (excluding the impact of dust) presents an approximately flat slope. For the large majority of our results we use the  luminosity of best-fitting stellar template at 1600\AA, however, in Section~\ref{sec:nebular_Acurve} we also make use of the {\it SFR} traced by the luminosity at 2800\AA. 

With the objective of deriving the {\it SFR} from the total IR luminosity ($L_{\mathrm{TIR}}$; frequently defined as the emission enclosed between 8 to 1000\,$\mu$m), \citetalias{2019ApJS..243...22B} distinguish between two cases: galaxies with only MIR detection (i.e., \textit{Spitzer}/MIPS 24 $\mu$m) and those with both MIR and FIR detections (i.e., \textit{Spitzer}/MIPS and \textit{Herschel}/PACS or SPIRE). \citetalias{2019ApJS..243...22B} obtain $L_{\mathrm{TIR}}$ for the former using the analytic conversion from MIPS 24$\mu$m luminosities to $L_{\mathrm{TIR}}$ published by \citet{2008ApJ...682..985W,2011ApJ...742...96W}. For the latter, \citetalias{2019ApJS..243...22B} compute $L_{\mathrm{TIR}}$ as the average value of the integrated IR luminosities obtained for the different best-fitting dust emission template of each library considered (i.e., \citealt{2001ApJ...556..562C}, \citealt{2002ApJ...576..159D}, \citealt{2009ApJ...692..556R}, and \citealt{2007ApJ...657..810D}). Only detections at a significance level $>$5$\sigma$ are considered.
The corresponding limiting fluxes of the {\it Spitzer}/MIPS 24\, and 70\,$\mu$m bands are 30 and 2500\,$\mu$Jy and. In the case of \textit{Herschel} data, the limiting fluxes for PACS (100 and 160\,$\mu$m) and SPIRE (250, 350 and 500\,$\mu$m)  bands are 1.6, 3.6, 9.0, 12.9, and 12.6\,$\mu$Jy, respectively.  
Consequently, this bolometric IR luminosity is transformed into values of {\it SFR} using the calibration by \citet[][transformed into a \protect\citealt{2003PASP..115..763C} IMF]{1998ARA&A..36..189K}: 
\begin{equation}
SFR_{\mathrm{TIR,\,K98}}/M_{\odot}yr^{-1} = 1.09 \times 10^{-10} L_\mathrm{TIR}/L_{\odot} \label{eq:SFRtir}
\end{equation}
\noindent The typical systematic offset between these two {\it SFR} estimations remain below 0.05~dex at $z$\,$<$\,3. 

\subsection{Best estimate of the total star-forming activity }\label{seq:SFRTOT}
\citetalias{2019ApJS..243...22B} also provide a {\it best estimate} of the total star-forming activity ({\it SFR}$_{\mathrm{TOT}}$) for the galaxies in the catalog. They derive it by applying different recipes depending on the information available for each galaxy. In the case of galaxies detected in the IR, they add the SF traced by the (unobscured) emission in the UV to that obscured probed by the dust emission in the IR ($SFR_{\mathrm{TOT}} = SFR_{\mathrm{TIR}} +  SFR^{\mathrm{obs}}_{\mathrm{UV}}$; \citealt{1998ARA&A..36..189K}, \citealt{2005ApJ...625...23B}). 
\begin{equation}
    SFR_{\mathrm{TOT}} = 1.09^{-10} [L_{\mathrm{IR}} + 3.3\times L_{\mathrm{UV}}]
\end{equation}
\noindent where the luminosities are in $L_{\odot}$. In this case, $L_{\mathrm{UV}}$ corresponds to the luminosity at 2800\AA. 

\begin{figure*}
\includegraphics[width=1\linewidth]{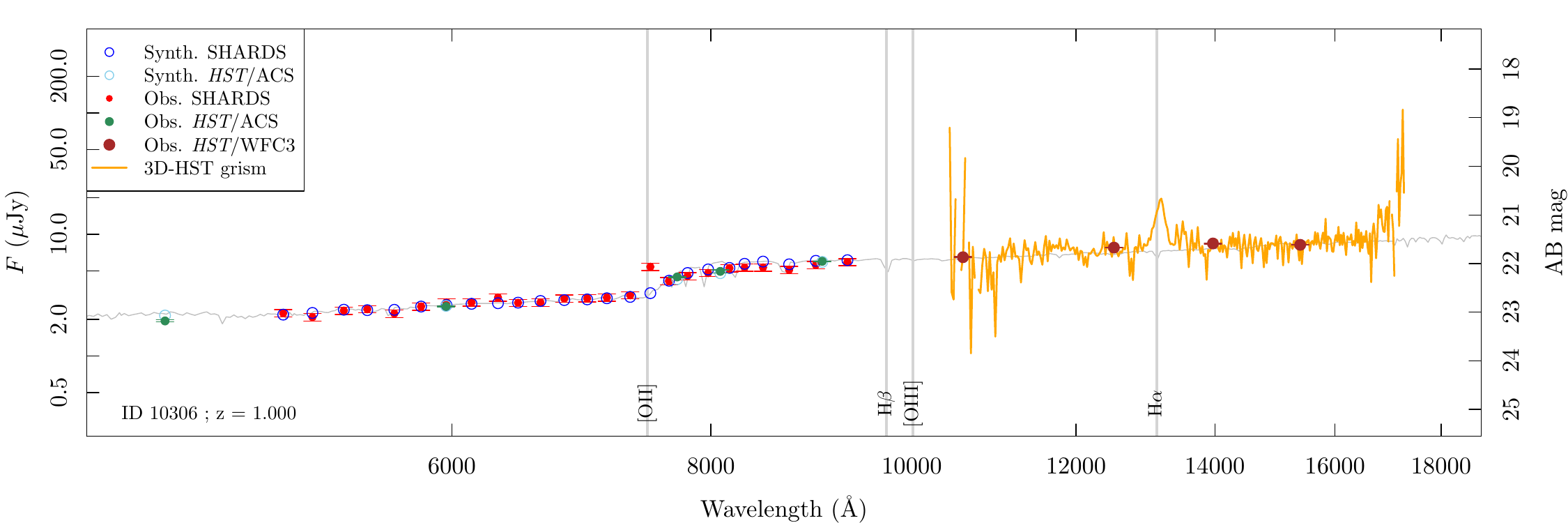}
\caption{SED-fitting example for a SHARDS \OII\,emitter. Filled (empty) symbols represent the observed (synthetic) \HST/ACS, \HST/WFC3 and SHARDS photometry, while the gray line shows the best-fitting continuum template. \HST\,grism data is also displayed (orange line). Vertical gray lines mark the position of some  relevant spectral features at the redshift of the galaxy ($z$$=$1.00). These features are:  \OII, \hb, [OIII]$\lambda$5007, and \hal. The stellar mass reported for this galaxy in the catalog by \citetalias{2019ApJS..243...22B} is \logM$=$9.87. This galaxy is detected in the IR and also presents \hal-emission in the {\it HST}/grism data. The luminosity of \OII\,measured on SHARDS data is (8.87$\pm$0.45)$\times$10$
^{41}$~erg\,s$
^{-1}$.} \label{fig:SED}
\end{figure*}

\section{Identification of \OII-emitters in SHARDS}\label{sec:oiiemitters}

Building on previous works (e.g., \citealt{2015ApJ...812..155C}, \citealt{2019A&A...621A..52L}) we have implemented a selection technique to identify ELGs in the SHARDS medium-band spectro-photometric data relying on a SED-fitting technique of the \HST/ACS, \HST/WFC3, and SHARDS available data. By comparing the observed flux densities in each SHARDS band ($F_{\mathrm{obs}}$) with the synthetic ones ($F_{\mathrm{syn}}$) obtained convolving the galaxy continuum emission best-fitting template with the response curve of the same filter, we can identify galaxies presenting an excess of $F_{\mathrm{obs}}$ due to ELs. Note that, as explained by \cite{2013ApJ...762...46P} and as a result of a construction instrumental feature, the position of the sources in the GTC/OSIRIS FoV determines the actual bandpass of each SHARDS filter. For this reason, the synthetic photometry is obtained using the passband \textit{seen} by each individual galaxy.

This methodology exploits the detailed information provided by SHARDS on the SED to obtain an accurate estimate of the continuum in each filter. This methodology is more robust against the impact of other ELs or abrupt changes in the SEDs (e.g., D4000 break) with respect to techniques relying in the interpolation of flux densities in adjacent filters or the use of a continuum broadband (e.g., \citealt{2008ApJS..176..301O,2008ApJ...677..169V,2011ApJ...740...47V, 2009MNRAS.398L..68S,2009MNRAS.398...75S,2012MNRAS.420.1926S, 2013MNRAS.428.1128S,2014MNRAS.440.2375M, 2015ApJ...812..155C}). 

\subsection{SHARDS photometry fine-tuning}
In order to guarantee the high quality of the EL flux measurements, we perform a fine-tuning calibration of the SHARDS photometry on a single galaxy basis. Our aim is to make the SHARDS photometry completely compatible with the {\it HST}/ACS data correcting hypothetical offsets between them. These offsets can arise for individual galaxies due to the different resolutions of the images on which their photometry is measured.  To tackle this issue, we perform a SED-fit of the {\it HST}/ACS and WFC3 photometry using \texttt{synthesizer} code (\citealt{2005ApJ...630...82P} \citealt{2008ApJ...675..234P}) with the same coverage of the parameter space as \citetalias{2019ApJS..243...22B} (see Section~\ref{sec:data}). We then calculate the convolution of the best-fitting SEDs with the SHARDS filters. We obtain the median and RMS of the ratio between the observed and synthetic SHARDS photometry of each galaxy. Finally, we apply this factor to the SHARDS photometry and propagate the photometric uncertainties. We do not update the photometry of galaxies with less than 5 SHARDS detections for which the impact of ELs could introduce undesirable offsets (i.e. over corrections). We also exclude from this procedure galaxies for which the RMS of the offset is larger than the 30\%. The median and percentiles P16$^{\mathrm{th}}$ and P84$^{\mathrm{th}}$ of the factor applied to shift SHARDS photometry are 0.95, 0.81, and 1.11. These numbers are obtained for the galaxies with magnitudes brighter than 26.5 in the {\it HST}/WFC3 F160W band. 

\subsection{Continuum estimate}\label{sec:sed}
We fit the available data in the UV-to-NIR regime (including the recalibrated SHARDS photometry) using the \texttt{synthesizer} code (\citealt{2005ApJ...630...82P}, \citealt{2008ApJ...675..234P}) in the overall same configuration as \citetalias{2019ApJS..243...22B} (see Section~\ref{sec:data}). In our case, we exclude the ELs from the SED templates. This is because our aim is to create synthetic photometry of the continuum to be able to identify observed flux excesses with respect to it. 
The redshift is fixed to the best redshift estimate given by \citetalias{2019ApJS..243...22B}, i.e., spectroscopic where available and photometric otherwise. The objective of the SED-fitting procedure is exclusively optimizing the measurements of the ELs through an accurate continuum estimate. Therefore, for the purpose of our work we use the physical properties obtained by \citetalias{2019ApJS..243...22B}. It is worth mentioning that the small offsets we apply to the SHARDS data do not lead to significant systematical differences with respect the work by \citetalias{2019ApJS..243...22B}. 
Figure~\ref{fig:SED} shows an example of the SED-fitting performed on the optical and NIR regimes of the emission of a galaxy in the catalog by \citetalias{2019ApJS..243...22B}. The continuum estimate for each band is obtained convolving each SHARDS filter with the best-fitting SED. The figure displays also the {\it HST} grism spectrum, which shows an emission in \hal.

\subsection{Detection of ELGs}\label{sec:detect}

\begin{figure*}
\includegraphics[width=1\linewidth]{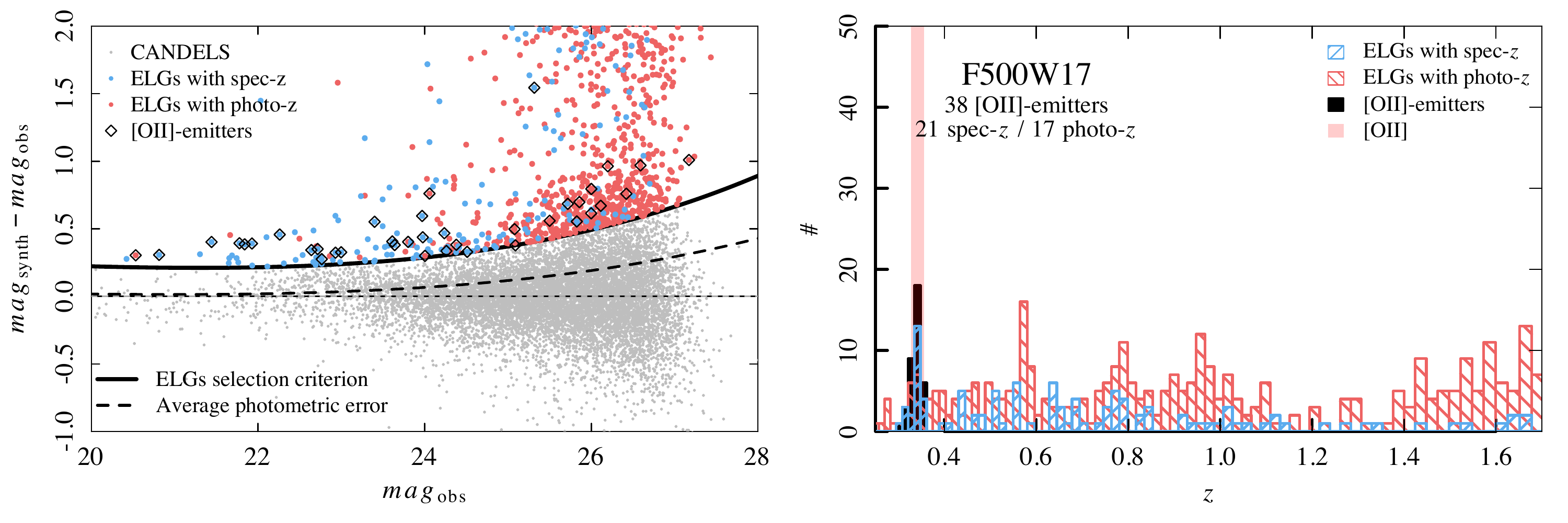}
\caption{{\it Left panel}: Color-magnitude diagram showing ELG candidates 
in SHARDS filter F500W17. Color is defined as the difference between the synthetic
and the observed magnitudes. The color threshold which defines the locus of ELGs is marked with a black continuous line. The color equal to zero is marked with a horizontal thin dashed black line. The dashed curve represent the average photometric error. {\it Right panel}: Redshift distribution 
of the ELG candidates throughout the redshift range in which \OII\, falls within the wavelength range over which SHARDS extends. The vertical coloured line marks the redshift that shifts the \OII\, line into the SHARDS F500W17 filter. The blue (red) histogram displays the distribution of galaxies with spectroscopic (photometric) redshifts. The black histogram shows the distribution of the 38 \OII-emitters detected in the F500W17 band. Appendix~\ref{A1} includes analogous figures for the rest of the SHARDS filters.} \label{fig:selection_one}
\end{figure*}

\begin{figure}
\includegraphics[width=1\linewidth]{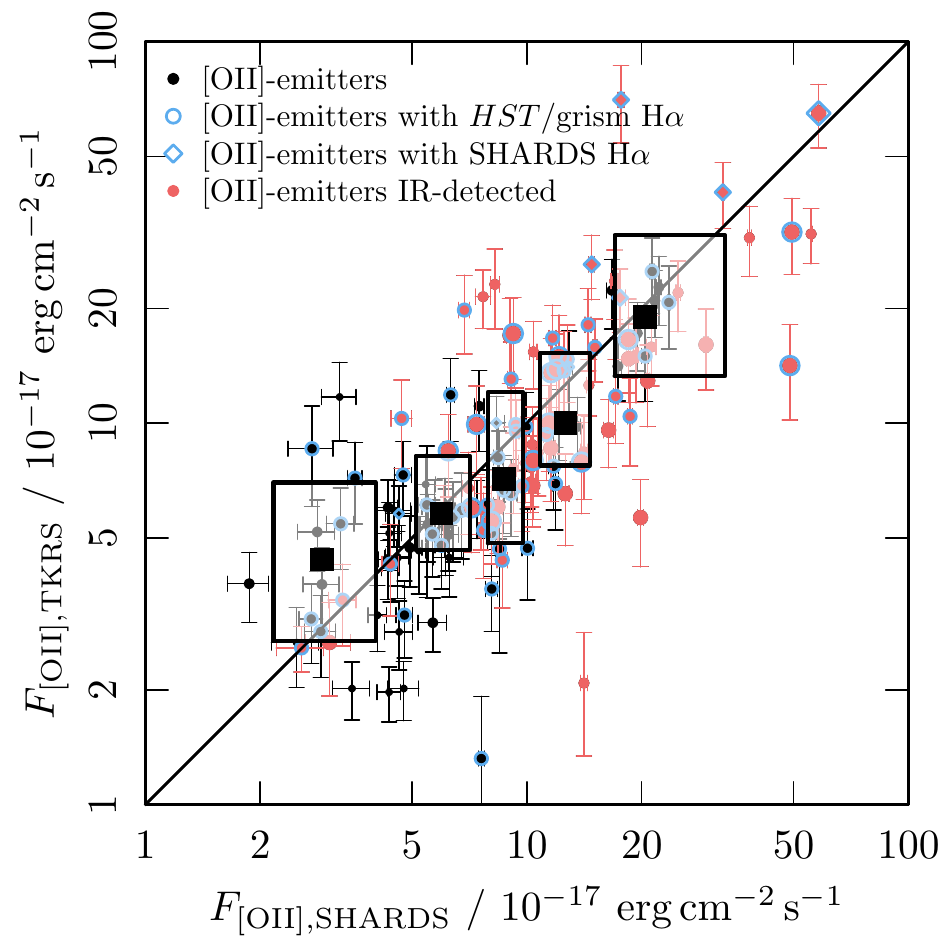}
\caption{Comparison between the \OII\,flux measurements performed on TKRS spectra and SHARDS spectrophotometric data. Black (red) symbols represent the \OII-emitters undetected (detected) in the FIR. Those \OII\,emitters with a \hal\,detection {in SHARDS or 3D-{\it HST} data}  are highlighted with a blue circle. The sizes of the symbols are mass dependent: small, medium, and large for log$_{10}M_*/M_{\odot}$<9, 9 < log$_{10}M_*/M_{\odot}$<10, and log$_{10}M_*/M_{\odot}$>10, respectively. The continuum black line represent the identity relation. Median values in four equally populated bins of SHARDS flux are represented using the median {(black filled squares)} and P16$^{\mathrm{th}}$ and P84$^{\mathrm{th}}$ quantiles {(black rectangles)}.}\label{fig:OII_spec_phot} 
\end{figure} 

We define our selection criteria in the color-magnitude plane, where the color is defined as the difference of the magnitudes corresponding to the $F_{\mathrm{obs}}$ and $F_{\mathrm{syn}}$ ($m_{\mathrm{obs}}$ and $m_{\mathrm{syn}}$, respectively), and it is represented as a function of the $m_{\mathrm{obs}}$. In Figure~\ref{fig:selection_one} (see also Appendix~\ref{A1}) we show the sketch of the technique used for the selection of the ELGs for the shortest wavelength SHARDS filter. 
 
 The color-magnitude diagram displays the data points in a \textit{trumpet}-like shape  distribution due to the fact that at fainter magnitudes, the photometric errors become larger, increasing the scatter. ELGs are located in the positive color \textit{locus} contributing to the positive wing of the distribution of color. Instead, the distribution of the data points in the negative color \textit{locus} is dominated by the contribution of the photometric errors and intrinsic differences between templates and photometry, rather than the presence of stellar absorptions. This is because the best-fitting spectral templates used to derive $m_{\mathrm{syn}}$ include absorption features (in contrast to ELs). In fact, our methodology allows us to detect and to measure ELs avoiding the effect of stellar absorption features. The negative color \textit{locus} can be used to estimate the typical scatter as a function of magnitude avoiding the impact of the presence of emitters. In practice, the intrinsic scatter of the color data points is evaluated in equally populated bins of $m_{\mathrm{obs}}$. The distance between P16$^{\mathrm{th}}$ and  P50$^{\mathrm{th}}$ (which defines the 1$\sigma$ scatter in the negative color $locus$) is then mirrored towards the positive values of color to find the P84$^{\mathrm{th}}$ of the intrinsic color distribution (i.e., which defines the 1$\sigma$ scatter in the positive color $locus$). Note that we use the median as the axis of this operation rather than the zero color value to be able to take into account slight offsets in the color distributions. Finally, the curve delimiting the intrinsic scatter at 1$\sigma$ is derived by fitting with a fourth order polynomial the values of these P84$^{\mathrm{th}}$ for every $m_{\mathrm{obs}}$ bin.  

The samples of ELG candidates are built by selecting in each SHARDS filter those galaxies with a detection at least at a 3$\sigma$ level for which the color is at least a factor 2 the scatter ($\sigma$) of the intrinsic color distribution for the given $m_{\mathrm{obs}}$ (see the left panel in Figure~\ref{fig:selection_one}). The use of this factor allows us to maximize the number of emitter candidates detected maintaining a high level of reliability of the final \OII-emitter samples, as defined in Section~\ref{sec:reliability}. In order to guarantee the significance of the differences between the observed and the synthetic magnitudes, we remove from the sample those galaxies for which their $m_{\mathrm{obs}}$-$m_{\mathrm{syn}}$ color is smaller than their photometric error. Combining the results in all filters, we detect 27090 emissions from 13183 ELGs. The ELGs detected in SHARDS represents $\sim$50\% of the WFC3/F160W selected parent catalog over the area covered by the SHARDS maps, including all redshifts.

\begin{table*}
\scriptsize
	\centering
	\caption{Samples of \OII-emitters. The table displays: (1) the name of the filter; (2) redshift at which \OII\, shifts to the average central wavelength of each SHARDS filter; (3) the number of \OII-emitters detected (including all masses), and (4-5) among them, those identified using the spec-$z$ and the photo-$z$, respectively; (6-7) the success rate and contamination; (8-12) the same quantities as in columns 3-7 for the {IR-detected \OII-emitters with log$_{10}$($M_{*}$/M$_{\odot}$)$>$9; (13-15) the number of  H$\alpha$-detected, {\it UVJ}-passive, and AGN candidates, respectively.}}
	\label{tab:counts}
	\begin{tabular}{cc|ccccc|cccccccc} 
		\hline
		SHARDS & $z_{\OII}$ & \multicolumn{5}{c}{\OII-emitters} & \multicolumn{8}{c}{{\OII-emitters [log$_{10}$($M_{*}$/M$_{\odot}$)$>$9 \& IR-detected]}} \\
		\cline{3-15} 
		filter & at $\lambda_{ \mathrm{central}}$ & All & spec-$z$ & photo-$z$ & {\it SR} & {\it C} & All & spec-$z$ & photo-$z$ & {\it SR} & {\it C} & With H$\alpha$ & $UVJ$-passive  & AGN\\
		(1) & (2) & (3) & (4) & (5) & (6) & (7) & (8) & (9) & (10) & (11) & (12) & (13) & (14) & (15) \\
		\hline
F500W17 & 0.34 &  38  &  21 & 17 &  0.76  &  0.07 &  5 &  5  &  0  &  0.80  &  0.00  &  5   &  0  &  0 \\
F517W17 & 0.39 &  66  &  29 & 37 &  0.91  &  0.05 &  4  &  4  &  0  &  1.00  &  0.00  &  4 &  0  &  4 \\
F534W17 & 0.43 &  154  &  80 & 74 &  0.71  &  0.16 &  23  &  21  &  2  &  0.73  &  0.15  &  12  &  2  &  9 \\ 
F551W17 & 0.48 &  222  &  121 & 101  &  0.92  &  0.12 &  42  &  40  &  2  &  0.97  &  0.03  &  5  &  1  &  1 \\
F568W17  & 0.52 &  240  &  122 & 118 & 0.92  &  0.06 &  37  &  37  &  0  &  0.85  &  0.04  &  0  &  1  &  0\\
F585W17 & 0.59 &  234  &  105 & 129 &  0.83  &  0.03 &  42  &  41  &  1  &  0.77  &  0.00  &  1  &  2  &  0\\
F602W17  & 0.62 &  123  &  44 & 79 &  0.70  &  0.07 &  17  &  17  &  0  &  0.60  &  0.00  &  0  &  1  &  0\\
F619W17  & 0.66 &  191  &  82 & 109 &  0.71  &  0.22 &  37  &  35  &  2  &  0.83  &  0.23 &  9  &  3  &  0\\
F636W17  & 0.71 &  126  &  53 & 73 &  0.97  &  0.06 &  13  &  12  &  1  &  1.00  &  0.00  &  9  &  0  &  0\\
F653W17 & 0.75 &  162  &  54 & 108  &  0.87  &  0.09 &  27  &  26  &  1  &  0.96  &  0.08  &   20  &  2  &  0\\
F670W17  & 0.80 &  209  &  88 & 121 &  0.79  &  0.00 &  36  &  35  &  1  &  0.80  &  0.00  &  27  &  1  &  0\\
F687W17  & 0.84 &  363  &  148 & 215 &  0.87  &  0.05 &  77  &  71  &  6  &  0.90  &  0.03  &   53  &  1  &  2\\
F704W17  & 0.89 &  238  &  63 & 175 &  0.88  &  0.08 &  35  &  29  &  6  &  1.00  &  0.00  &  24  &  3  &  0\\
F721W17 & 0.93 &  286  &  125 & 161 &  0.90  &  0.03 &  73  &  63  &  10  &  0.92  &  0.04  &  55  &  6  &  0\\
F738W17  & 0.98 &  272  &  81 & 191 &  0.75  &  0.29 &  39  &  37  &  2  &  0.89  &  0.15  &  23  &  0  &  0\\
F755W17 & 1.02 &  266  &  131 & 135 & 0.86  &  0.08 &  60  &  52  &  8  &  0.93  &  0.05  &   51  &  5  &  0\\
F772W17  & 1.07 &  94  &  26 & 68  &  1.00  &  0.11 &  12  &  8  &  4  &  1.00  &  0.00  & 9  &  1  &  0\\
F789W17 & 1.12 &  74  &  13 & 61 &  0.86  &  0.00 &  8  &  6  &  2  &  0.50  &  0.00  &   5  &  2  &  0\\
F806W17 & 1.16 &  173  &  52 & 121 &  0.84  &  0.14 &  35  &  29  &  6  &  0.95  &  0.17  &  23  &  3  &  3\\
F823W17 & 1.21 &  155  &  45 & 110 &  0.92  &  0.03 &  20  &  14  &  6  &  1.00  &  0.07  &  16  &  3  &  0\\
F840W17  & 1.25 &  174  &  52 & 122 &  0.94  &  0.11 &  38  &  26  &  12  &  0.95  &  0.05  &  27  &  1  &  0\\
F857W17 & 1.30 &  120  &  17 & 103  &  1.00  &  0.20 &  11  &  8  &  3  &  1.00  &  0.17   &  6  &  0  &  0\\
F883W35  & 1.37 & 169  &  42 & 127 &  0.88  &  0.03 &  30  &  18  &  12  &   0.93  &  0.07  &  23  &  0  &  0\\
F913W25  & 1.45 &  163  &  27 & 136 &  0.85  &  0.11 &  21  &  8  &  13  &  1.00  &  0.00 &  11  &  0  &  1\\
F941W33  & 1.52 & 141  &  31 & 110 &  0.92  &  0.04 &  21  &  11  &  10  &  1.00  &  0.00 &  12  &  0  &  0\\ 
		\hline
Total & & 4455  &  1653 & 2802 & & &  763  &  653  &  110  &  &  &  430  &  38  &  20\\
\hline
	\end{tabular}
\end{table*}

\subsection{Identification of \OII-emitters}\label{sec:ident}
We identify \OII-emitters among the previously described sample of ELG candidates using their spec-$z$ and photo-$z$. As mentioned in Section~\ref{sec:data_z}, SHARDS data enable deriving extremely high quality photo-$z$'s, which translates in the capability of building robust samples of ELGs.
The process to identify \OII-emitters consists in finding for each filter those ELGs located within the range of redshift (spec-$z$ when possible and photo-$z$ otherwise) that would shift the rest-frame \OII\,lines into such band. In practice, the selection ranges are simply defined by the redshift windows in which the \OII\,falls within the full width half maximum (FWHM) of each SHARDS medium-band filter. {We use an analogous approach to identify galaxies with an \hal\, detection in SHARDS data among the sample of \OII-emitters.} 

Table~\ref{tab:counts} shows the number of \OII-emitters identified in each filter. Note that the number counts in the filters F687W17 and F823W17 exceed those obtained by \citet[][285 and 142, respectively]{2015ApJ...812..155C} based on an alternative methodology in which the continuum estimate relied on the interpolation of the flux densities of adjacent SHARDS bands.

We find a total of 4455 \OII-emitters. Among them, 1653 are spectroscopically confirmed and 2802 are selected using their high quality photo-$z$. Figure~\ref{fig:selection_one} shows the redshift distribution of the emitters in the shortest wavelength SHARDS filter.  Appendix~\ref{A1} contains the same plot for the rest of the SHARDS filters. The average percentage of emitters selected based on their photo-$z$ (spec-$z$) increases (decreases) with redshift from approximately 50\% (50\%) to 70\% (30\%). 

\subsection{Reliability of the sample of \OII-emitters} \label{sec:reliability}
We assess the level of reliability and purity of the sample of \OII-emitters using two quantities similar to those defined by \citet{2015ApJ...812..155C}: the \textit{success rate} ({\it SR}) and the \textit{contamination} ({\it C}). These quantities are calculated using exclusively the galaxies with an available spec-$z$.
The first step to compute {\it SR} and {\it C} is performing an additional identification of \OII\,-emitters among the sources with an available spec-$z$ following the same procedure outlined in Section~\ref{sec:ident}, but using their corresponding photo-$z$. 
{\it SR} is defined as the fraction of galaxies identified as \OII-emitters in a given filter using their spec-$z$\,($N_{\OII, \mathrm{spec}}$) that are also identified as \OII-emitters  using their photo-$z$ ($N^{\mathrm{conf}}_{\OII, \mathrm{phot}}$):  
\begin{equation}
SR=N^{\mathrm{conf}}_{\OII, \mathrm{phot}}/N_{\OII, \mathrm{spec}}
\end{equation}
The {\it SR} is related to the ability of our method (based on medium-band SEDs and photo-$z$ determination) to identify {\it bona fide} ELGs. {For 10 SHARDS filters, we recover virtually all confirmed emitters (SR $\geq$90\%). We are able to identify more than 80\% (70\%) of the confirmed emitters in 19 (all 25) filters. } Table~\ref{tab:counts} gives the SR obtained for each filter. 

{\it C} is the fraction of galaxies with an available spec-$z$ and that are selected as \OII-emitters when considering their phot-$z$ (independently of the value of their spec-$z$; N$_{\OII, \mathrm{phot}}$), that are not \OII-emitters if their spec-$z$ are used. We can calculate this following the expression:
\begin{equation}
C=1-N^{\mathrm{conf}}_{\OII, \mathrm{phot}}/N_{\OII, \mathrm{phot}}
\end{equation}
This number gives an estimate of the fraction of contaminants expected in the final samples. The contamination found for each filter is shown in Table~\ref{tab:counts}. {For 16 out of 25 filters we obtain $C\leq$10\%. The contamination remains below 20\% (30\%) for 22 (all 25) filters.} 
Our new analysis provides similar numbers to those found by \citet{2015ApJ...812..155C} for F687W17 and F823W17. The sample of \OII-emitters presents a small fraction of missed and spurious sources, probably linked to photometric and redshift uncertainties. 

\section{Measurement of EL fluxes in SHARDS data} \label{sec:flux}
We have measured line fluxes ($F$) by applying the following definition (see, e.g., \citealt{2008ApJ...677..169V}, \citealt{2012MNRAS.420.1926S}):
\begin{equation}
F = (F^i_{\mathrm{obs}}-F^i_{\mathrm{syn}})\times \Delta^i
\end{equation}
where $i$ indicates the different SHARDS filters, and $\Delta^i$ is the FWHM of the $i$-th filter. We then obtain the corresponding luminosities considering the luminosity distance for the redshift of each galaxy. As it is customary, we propagate the errors of the observed photometry to derive the uncertainties of the measured fluxes and luminosities.  

We assess the reliability of our \OII~measurements based on SHARDS spectrophotometric data by comparing these values to those measured on available spectroscopic data for a subsample of 148 \OII-emitters. In particular, we use data from the Team Keck Redshift Survey (TKRS; \citealt{2004AJ....127.3121W}). Figure~\ref{fig:OII_spec_phot} shows an overall good agreement between the two measurements with a wide scatter (normalized median absolute deviation 40\%, \citealt{1983ured.book.....H}). Fluxes measured in the TKRS spectroscopy are on average 4\% larger than the fluxes measured in SHARDS. No systematic differences are observed for \OII\, emitters with or without an available \hal\, measurement {(in SHARDS or 3D-{\it HST} data)} or a detection in the IR ({\it Spitzer}/MIPS 24$\mu$m and/or {\it Herschel}).

\section{The final sample of SHARDS IR-detected \OII-emitters}\label{sec:physprop}
\begin{figure}
\includegraphics[width=0.99\linewidth]{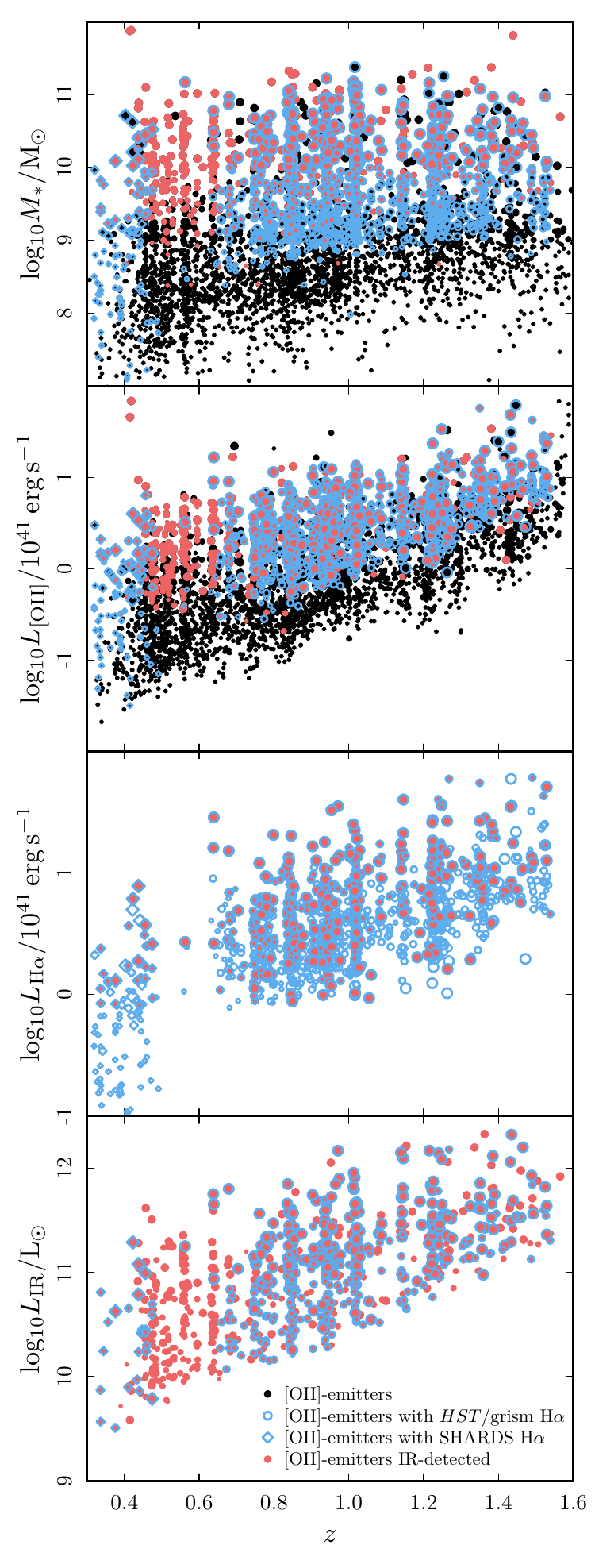}
\vspace{-0.7cm}
\caption{Stellar masses, \OII\, luminosities, \hal\,luminosities (from {\it HST}/grism and SHARDS data; only \hal\, detections at SNR$>$3), and total IR luminosities (see Section~\ref{seq:SFRCONT}) of the sample of SHARDS \OII-emitters as a function of redshift. The \hal\,luminosities are corrected for [NII]$\lambda\lambda$6568.1,6583.6 contamination (see Section~\ref{seq:SFREL}). The size of the data points scales with the stellar mass: small, medium, and large for log$_{10}M_*/\mathrm{M}_{\odot}$$<$9, 9$<$ log$_{10}M_*/\mathrm{M}_{\odot}$$<$10, and log$_{10}M_*/\mathrm{M}_{\odot}$$>$10, respectively.}\label{fig:m_L_z}
\end{figure}

{We focus our analysis on the subsample of \OII-emitters with stellar mass larger than the mass completeness limit at $z$$\sim$1.5 [log$_{10}$($M_{*}$/M$_{\odot}$)$>$\,9] and a MIR or FIR detection at a significance level $>$3$\sigma$. Table~\ref{tab:counts} reports the number counts and the reliability evaluation of the final sample in each filter. 
Figure~\ref{fig:m_L_z} displays the distribution with redshift of the stellar mass, luminosities of \OII\,and \hal\,emission lines, and the total IR luminosity of the full sample of SHARDS \OII-emitters. The upper panel shows the power of the SHARDS survey in identifying low-mass ELGs out to redshifts as high as 1.5. Future works will explore this low-mass population in detail. Section~\ref{sec:uvj} and Section~\ref{sec:AGN} asses the  contamination from quiescent galaxies and active galactic nucleus (AGN), respectively. In  Section~\ref{sec:fsample}, we briefly characterize the final sample in the context of its parent sample of \OII-emitters and the general population of SFGs throughout the redshift range of interest.}

\subsection{$UVJ$-diagram}\label{sec:uvj}
In literature, different techniques are used to build samples
of SFGs. Among them, and besides the identification of ELs, we find
certain color-color criteria. For instance, 
the rest-frame $U-V$ versus $V-J$ color-color space ($UVJ$-diagram) allows to select relatively pure samples of either quiescent or SFGs
(e.g. \citealt{2007ApJ...655...51W}, \citealt{2011ApJ...739...24B},
\citealt{2012ApJ...754L..29W,2015ApJ...811L..12W}). In
particular, we identify passive galaxies following the
recipes by \citet{2009ApJ...691.1879W}:
\begin{equation}
\begin{aligned}
&\left.
\begin{array}{l}
U-V>0.88\times(V-J)+0.69\\ 
U-V>1.3\\ 
V-J<1.6 
\end{array}
\right\} \,\,\,\,\,\,\, \mathrm{at}\,\,  0.0<z<0.5 \\
&\left.
\begin{array}{l}
U-V>0.88\times(V-J)+0.59\\ 
U-V>1.3\\ 
V-J<1.6
\end{array}
\right\} \,\,\,\,\,\,\, \mathrm{at}\,\,  0.5<z<1.0 \\
&\left.
\begin{array}{l}
U-V>0.88\times(V-J)+0.49\\
U-V>1.3\\
V-J<1.6
\end{array}
\right\} \,\,\,\,\,\,\, \mathrm{at}\,\, 1.0<z<1.5
\end{aligned}
\end{equation}
\noindent Galaxies that are not classified as passive are considered SFGs.

\begin{figure}
\includegraphics[width=\linewidth]{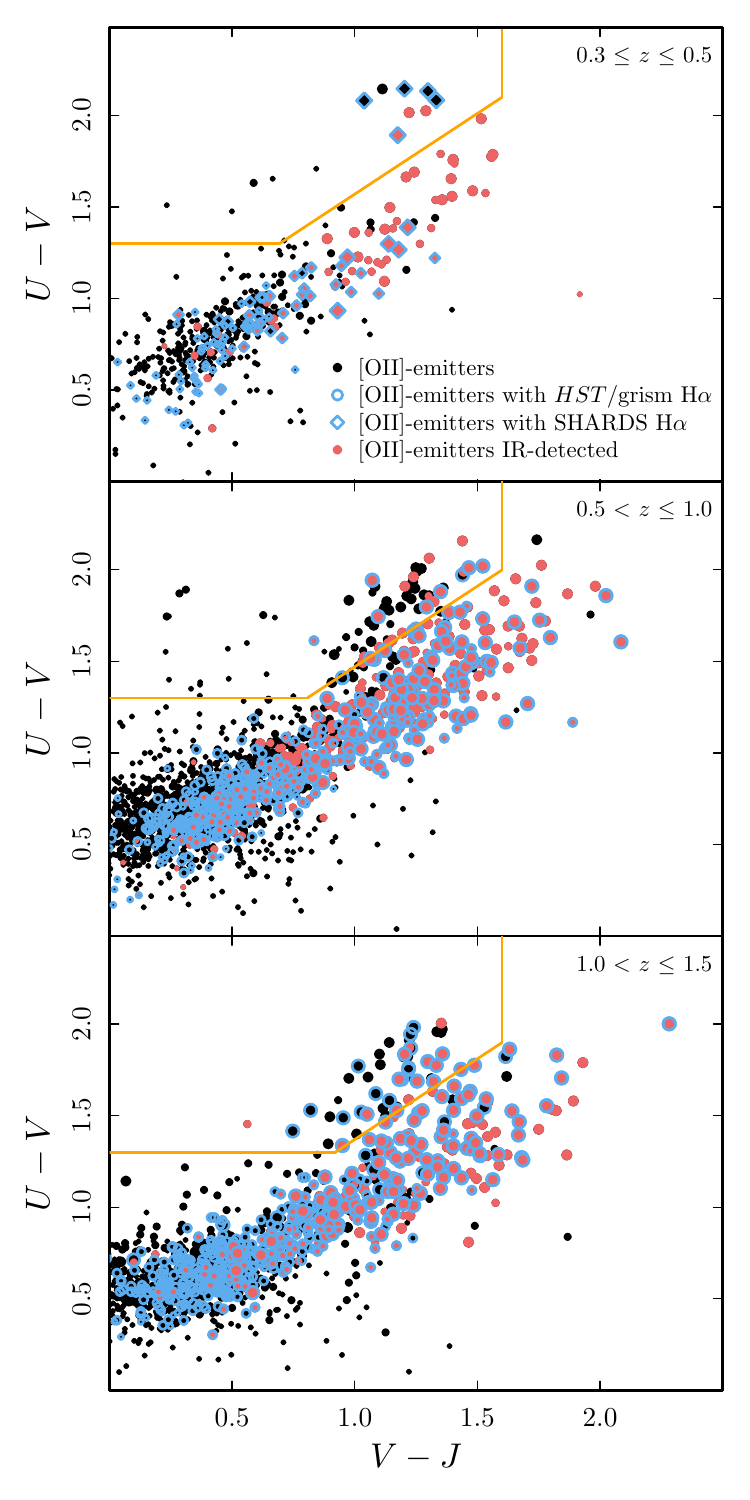}
\caption{Rest-frame {\it U-V} and {\it V-J} colors in three increasing redshift bins from top to bottom: 
{0.3$\leq$$z$$\leq$0.5, 0.5$<$$z$$\leq$1.0, and 1.0$<$$z$$\leq$1.5}. Orange lines mark the boundaries defined by \protect\citet{2009ApJ...691.1879W}
to distinguish between quiescent and SFGs in the corresponding redshift bins. Symbols as in Figure~\protect\ref{fig:OII_spec_phot}.}\label{fig:UVJ}
\end{figure}

\begin{figure}
\includegraphics[width=\linewidth]{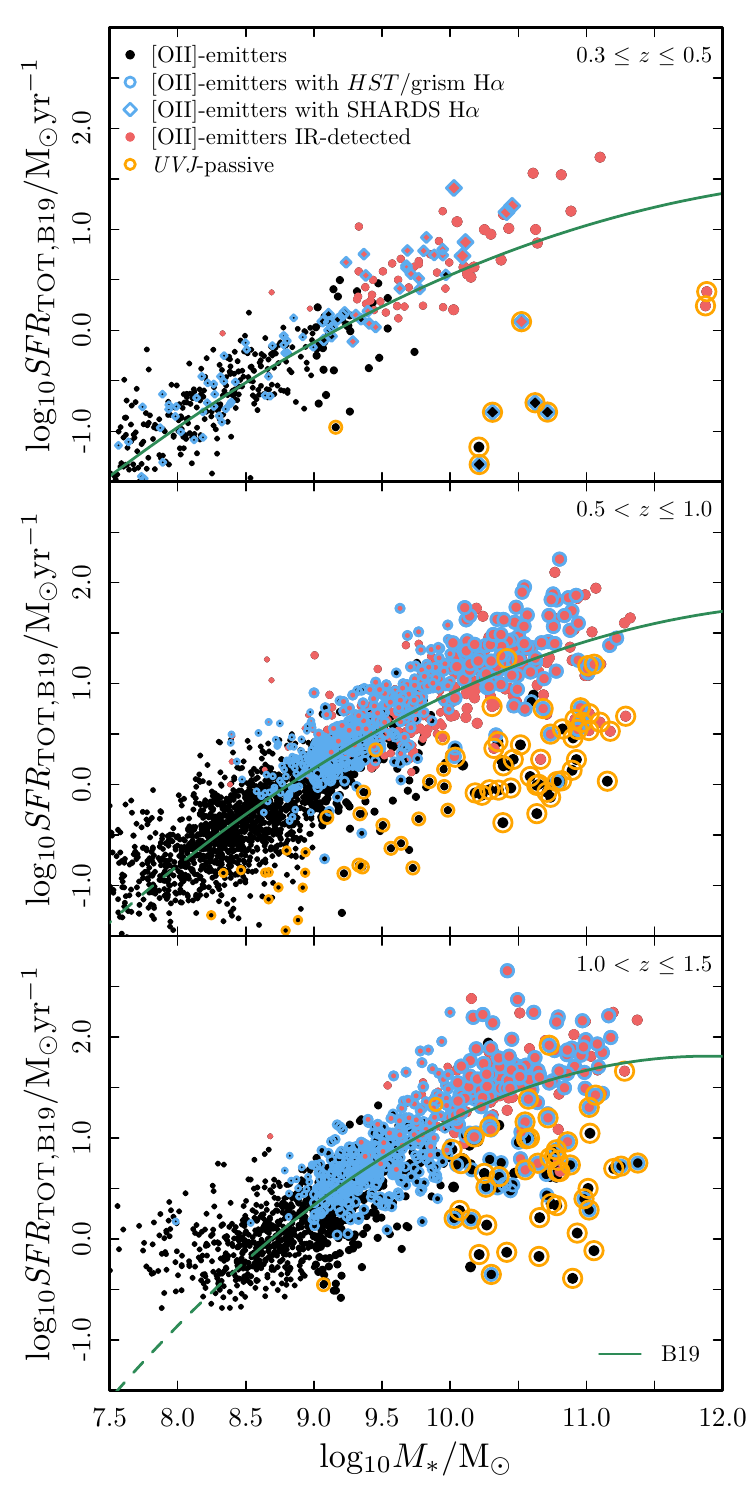}
\caption{
SFR$_{\mathrm{TOT}}$ \textit{versus} $M_{*}$ relation for the \OII-emitters split up in three increasing $z$ bins (top, middle, and bottom panels). The {\it SFR}$_{\mathrm{TOT}}$ refers to the {\it SFR}$_{\mathrm{TIR}}$\,$+$\,SFR$_{\mathrm{UV}}$ for those galaxies IR-detected, and {\it SFR}$_{\mathrm{UV,corr}}$ otherwise (see \citetalias{2019ApJS..243...22B} for details). Symbols as in 
Figure~\protect\ref{fig:OII_spec_phot}. {\it UVJ}-passive galaxies are highlighted with an orange circle. The green lines represent the main sequences fitted by \citetalias{2019ApJS..243...22B} to samples of galaxies selected as star-forming in the {\it UVJ}-diagram. The relations are extended beyond the stellar mass completeness level using dashed lines.}\label{fig:m_s}
\end{figure}

We explore the distribution of \OII-emitters in the $UVJ$-diagrams displayed in Figure~\ref{fig:UVJ}. Up to 94\% of the sample (95\%, 94\%, and 94\% in each of the three increasing redshift bins) is characterized by colors typical of SFGs. This result confirms the ability of our technique to exploit the SHARDS high quality data in order to identify highly pure ELG samples. Among those \OII-emitters in the locus of passive galaxies we find 33\%, 33\%, and 29\% with emission in the MIR or FIR, and 50\%, 13\%, and 48\% with \hal\,detection in \textit{HST} data or SHARDS bands. We note that below $z\sim$0.5 \hal\,falls in the SHARDS wavelength range. The cross contamination between the \textit{locii} of SFGs and passive galaxies in the $UVJ$-diagram has been found in previous studies (e.g., \citealt{2015ApJ...812..155C}, \citealt{2016MNRAS.457.3743D}). We find 117 passive galaxies out of the 1965 \OII-emitters with log$_{10}$($M_{*}$/M$_{\odot}$)$>$9.
Only 5\% of the IR-detected \OII-emitters with \logM$>$9 qualify as {\it UVJ}-passive systems. Figure~\ref{fig:m_s} displays the distribution of the total sample of \OII-emitters on the stellar mass {\it vs} {\it SFR} plane for the same three redshift bins used in Figure~\ref{fig:UVJ}. We can see that the {\it UVJ}-passive galaxies are located below the main sequence, as expected. We exclude these galaxies from the analysis. 

\subsection{AGN contamination}\label{sec:AGN}
Our methodology to select ELGs makes no difference between pure SFGs, AGN, and composite systems. We use the 2~Ms~CDFN X-ray catalogs published by \citet{2016ApJS..224...15X} to explore the contamination of the two latter type of systems. We find 85 X-ray counterparts of detected \OII-emitters with log$_{10}$($M_{*}$/M$_{\odot}$)$>$9 within a 2\,arcsec search radius. Among them, 20 {(14 IR-detected)} qualify as X-ray emitters with a $L_{\mathrm{X}}$$=$$10^{42}$~erg\,s$^{-1}$. This is the luminosity criteria normally used to identify AGN dominated systems (e.g., \citealt{2015ApJ...812..155C}). 
We also find 14 sources {(9 IR-detected)} with a counterpart in the catalog of variable sources published by \citet{2010ApJ...723..737V}, Two galaxies are included in both AGN candidate types. The fraction of AGN among the \OII-emitters [log$_{10}$($M_{*}$/M$_{\odot}$)$>$9] is 2\%. This result is consistent with the 1\%-2\% estimated by previous works in the same redshift range (e.g., \citealt{2009ApJ...701...86Z}, \citealt{2015ApJ...812..155C}). {In the case of the final sample of IR-detected \OII-emitters with log$_{10}$($M_{*}$/M$_{\odot}$)$>$9, the AGN fraction is 3\%. We exclude both the variable sources and the X-ray luminous systems from our sample of \OII-emitters. Two of the excluded objects are also {\it UVJ}-passive systems, among which only one is IR-detected.}

\subsection{The final sample} \label{sec:fsample}

\begin{table*}
\scriptsize
	\centering
	\caption{{Average properties of the} samples of IR-detected \OII-emitters with log$_{10}$($M_{*}$/M$_{\odot}$)$>$9 used to derive the results of this work. The table displays: (1) brief description of the subsample; (2) number of galaxies; the distribution of (3) redshift, (4) stellar mass, (5) UV attenuation (Equation~\ref{eq:A}), (6) best total {\it SFR} estimate by \citetalias{2019ApJS..243...22B}, (7-8) {\it SFR}s as traced by the IR and the UV (1600\,\AA), (9-11) {\it SFR} obtained from \OII\, following the calibration by \citetalias{2004AJ....127.2002K} and \citetalias{1998ARA&A..36..189K}, and from \hal\,through the calibration by \citetalias{1998ARA&A..36..189K}, (12) distance to the MS. The distributions are described with the median and the percentiles P16$^{th}$ and P84$^{th}$.}
	\label{tab:samples}
	\begin{tabular}{l|ccccccccccc} 
		\hline
		Sample of & \# & $z$ & log$_{10}M_{*}$ &
		$A_{\mathrm{UV}}$ &
		$SFR_{\mathrm{TOT}}$ & $SFR_{\mathrm{IR}}$ & 
		$SFR_{\mathrm{UV}}$ &
		$SFR_{\mathrm{[OII]K04}}$ & $SFR_{\mathrm{[OII]K98}}$ & 
		$SFR_{\mathrm{H}\alpha\mathrm{K98}}$ &
		$\Delta$MS \\
		\OII-emitters &  & & (M$_{\odot}$) &  & (M$_{\odot}$yr$^{-1}$) & (M$_{\odot}$yr$^{-1}$) & (M$_{\odot}$yr$^{-1}$) & (M$_{\odot}$yr$^{-1}$) & (M$_{\odot}$yr$^{-1}$) & (M$_{\odot}$yr$^{-1}$) & (dex)\\
		(1) & (2) & (3) & (4) & (5) & (6) & (7) & (8) & (9) & (10) & (11) & (12) \\
		\hline
		 IR-detected  & 706  &  0.87 $^{ 1.22 }_{ 0.56 }$ & 10.00 $^{ 10.55 }_{ 9.58 }$ &  3.05 $^{ 4.73 }_{ 1.79 }$ &  13.79 $^{ 40.57 }_{ 4.68 }$ &  11.27 $^{ 35.15 }_{ 3.00 }$ &  0.80 $^{ 2.37 }_{ 0.21 }$ &  1.08 $^{ 2.67 }_{ 0.42 }$ &  2.31 $^{ 5.68 }_{ 0.89 }$ &  &  0.15 $^{ 0.40 }_{ -0.10 }$ \\
		 IR-detected \& H$\alpha$ &   {396}  &  0.94 $^{ 1.24 }_{ 0.78 }$ & {10.00} $^{ 10.56 }_{ 9.63 }$ &  3.09 $^{ 4.68 }_{ 1.85 }$ &  {16.62} $^{ 43.88 }_{ 6.64 }$ &  {14.26} $^{ 40.38 }_{ 4.34 }$ &  {1.10} $^{ 2.64 }_{ 0.30 }$ &  {1.35} $^{ 3.08 }_{ 0.52 }$ &  {2.88} $^{ 6.56 }_{ 1.10 }$ &  {2.66 $^{ 5.93 }_{ 1.11 }$} &  {0.20} $^{ 0.45 }_{ 0.00 }$ \\
		\hline
	\end{tabular}
\end{table*}

\begin{figure*}
\includegraphics[width=\linewidth]{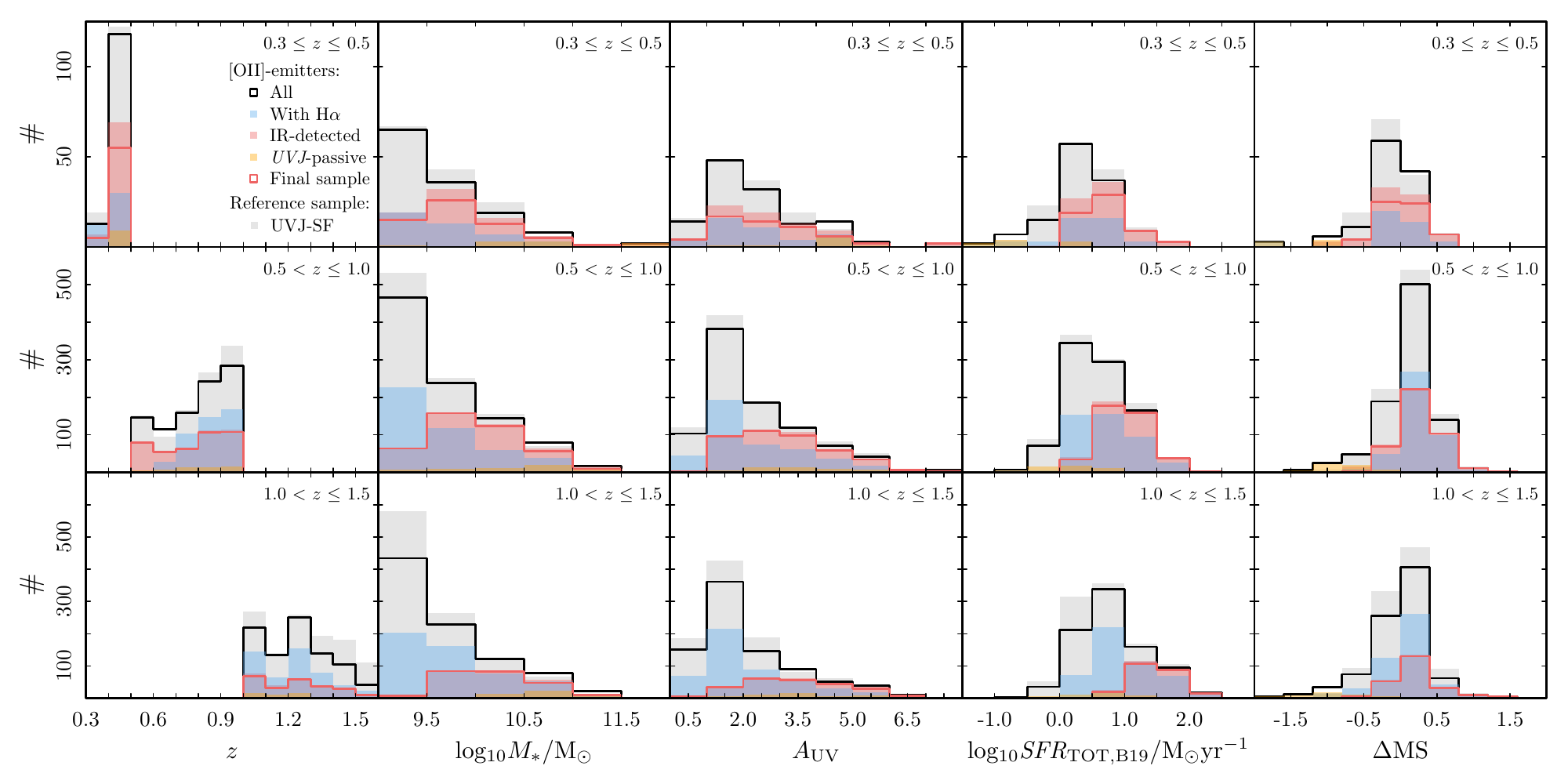}
\caption{{From left to right the distribution of redshift, stellar mass, UV attenuation (see Section~\ref{sec:attenuation}), {\it SFR}$_{\mathrm{TOT}}$, and distance to the main sequence for relevant subsamples of \OII-emitters with \logM$>$9 in the same redshift bins as in Figure~\ref{fig:m_s} and Figure~\ref{fig:UVJ}. For comparison, we use a {\it UVJ} selected sample of SFGs with \logM$>$9 and the same redshift distribution as the parent sample of [OII]-emitters.    }}\label{fig:histo}
\end{figure*}

{We end up with 706 IR-detected \OII-emitters with \logM$>$9 expanding over the redshift range  0.31$\leq$$z$$\leq$1.57. For some of the results we present in this work we make use of a subsample of 396 systems with a detection of \hal\, in either SHARDS or {\it HST}/grism data. 
Table~\ref{tab:counts} presents the number counts, and the reliability assessment of the final sample per filter. The table also shows the number of AGN candidates and {\it UVJ}-passive galaxies excluded. Finally, Table~\ref{tab:samples} gives a summary of the average physical properties of these two samples. }

{In order to further put the final sample of \OII-emitters in the context of the general SFG population, we use a reference sample of galaxies extracted from the same SHARDS/CANDELS catalog. This reference sample includes only galaxies that qualify as star-forming for their {\it UVJ} colors, and that are detected ($>$ 3$\sigma$) in the SHARDS filter into which their redshift shifts the restframe 3727\AA\,wavelength (i.e. same redshift range of the \OII-emitters). AGN candidates are excluded.}

{Figure~\ref{fig:histo} displays the distribution of some relevant properties of the aforementioned samples of \OII-emitters and the {\it UVJ}-SFG reference sample in three redshift bins. The figure reports the redshift, stellar mass, UV attenuation (Section~\ref{sec:attenuation}), total estimates of SF, and the distance to the main sequence by \citetalias{2019ApJS..243...22B} ($\Delta$MS~=~log$_{10}$$SFR_{\mathrm{TOT}}$-log$_{10}$$SFR_{\mathrm{MS}}$; see Figure~\ref{fig:m_s}). We clarify that $\Delta$MS is calculated along constant values of mass (i.e. not perpendicularly to the MS). The more positive (negative) $\Delta$MS is, the stronger (weaker) is the burst. The figure evidences the similarities between the samples of SFGs selected by the detection of \OII\, in SHARDS and {\it UVJ}-color criteria. Furthermore, the distributions show that, as a result of the selection function, there is a dependence of the average properties of the final sample of IR-detected \OII-emitters with redshift. Only $\Delta$MS appears statistically invariant, as it is calculated using a redshift dependent MS.} 

\section{The \lowercase{{\it f}}-factor for SHARDS \OII-emitters}\label{sec:ffactor} 
{We quantify the differential reddening by comparing  a  robust estimate of the total star formation activity ($SFR_{\mathrm{TOT}}$ by \citetalias{2019ApJS..243...22B}) with the {\it SFR} obtained using the \OII\, and \hal\, luminosities corrected for dust attenuation. This method has been used previously in literature (e.g., \citealt{2006ApJ...647..128E}, \citealt{2013ApJ...777L...8K}, \citealt{2015A&A...582A..80T}, \citealt{2016A&A...586A..83P}). In the following sections, we describe the recipes adopted to convert \OII\, and \hal\, luminosities into {\it SFR}, the framework of the methodology, and the results.}

\subsection{{\it SFR} from \OII\,and \hal\,luminosities} \label{seq:SFREL}
One of the most frequently used calibrations to transform \OII\,luminosity into a {\it SFR}, is the one published by \citet{1998ARA&A..36..189K}, {transformed into a \citet{2003PASP..115..763C} IMF}. This recipe uses the {\it SFR} calibration of the \hal\,luminosity reported in the same article, and assumes an average \OII/\hal\,ratio (0.57$\pm$0.06) not corrected for dust attenuation. 
\begin{equation}
SFR_{\mathrm{\OII, K98}}/M_{\odot}yr^{-1} =\, \,8.2 \times 10^{-42} L_\mathrm{\OII}/\mathrm{erg\,s}^{-1}\label{eq:SFRoii04}
\end{equation}
For a detailed explanation of the assumptions and important issues implicit in the \citetalias{1998ARA&A..36..189K} calibration, we refer the reader to \citetalias{2004AJ....127.2002K} work. 

We decide to use also the  alternative calibration published by \citetalias{2004AJ....127.2002K} {(transformed into a \citealt{2003PASP..115..763C} IMF)}, which differs from that by \citetalias{1998ARA&A..36..189K} in the average \OII/\hal\,ratio considered (1.2$\pm$0.3), which is corrected for dust-attenuation:
\begin{equation}
SFR_{\mathrm{\OII,\,K04}}/M_{\odot}yr^{-1} =\, \, 3.87 \times 10^{-42} L_\mathrm{\OII}/\mathrm{erg\,s}^{-1}\label{eq:SFRoiik98}
\end{equation}

We do not make use of the optional correction for metallicity provided by \citetalias{2004AJ....127.2002K} due to the {lack of information to obtain the oxygen abundance through the diagnostics for which they provide parametrizations}. Several other calibrations can be found in literature to transform \OII\,luminosities into {\it SFR}s (e.g., \citealt{1989AJ.....97..700G}, \citealt{1998ApJ...504..622H}, \citealt{2001ApJ...551..825J}, \citealt{2002MNRAS.332..283R},  \citealt{2003ASPC..297..191A}, \citealt{2006ApJ...648..281Y},  \citealt{2006ApJ...642..775M}, \citealt{2013MNRAS.430.1042H}), however analysing the differences between them is beyond the scope of this work. 

For \hal, we use the calibration by \citetalias{1998ARA&A..36..189K}, {which we modify to make it consistent with a \citet{2003PASP..115..763C} IMF}: 
\begin{equation}
SFR_{\mathrm{H}\alpha,\,K98}/M_{\odot}yr^{-1} =\, \,4.8 \times 10^{-42} L_{\mathrm{H}\alpha}/\mathrm{erg\,s}^{-1} \label{eq:SFRha}
\end{equation}
Figure~\ref{fig:m_L_z} displays the redshift distribution of the \OII\,and \hal\,luminosities measured for all the SHARDS \OII-emitters. We note that the \hal\,fluxes used here are both those from 3D-$HST$ included in the \citetalias{2019ApJS..243...22B} catalog (see also \citealt{2016ApJS..225...27M}) and those measured on SHARDS data. {In both cases, they are} corrected for [NII]$\lambda\lambda$6568.1,6583.6 contamination. 
{We apply a stellar mass and redshift dependent correction to account for the significant metallicity dependence on these two parameters. We use the parametrization of the mass-metallicity relation by \citet{2014ApJ...789L..40W} to derive oxygen abundances given a stellar mass and a redshift. Then, we use the linear metallicity calibration by \citet{2004MNRAS.348L..59P} to convert oxygen abundances into [NII]/H$\alpha$ fractions, from which we derive a correction factor.}

\begin{figure*}
\includegraphics[page=1,width=\linewidth]{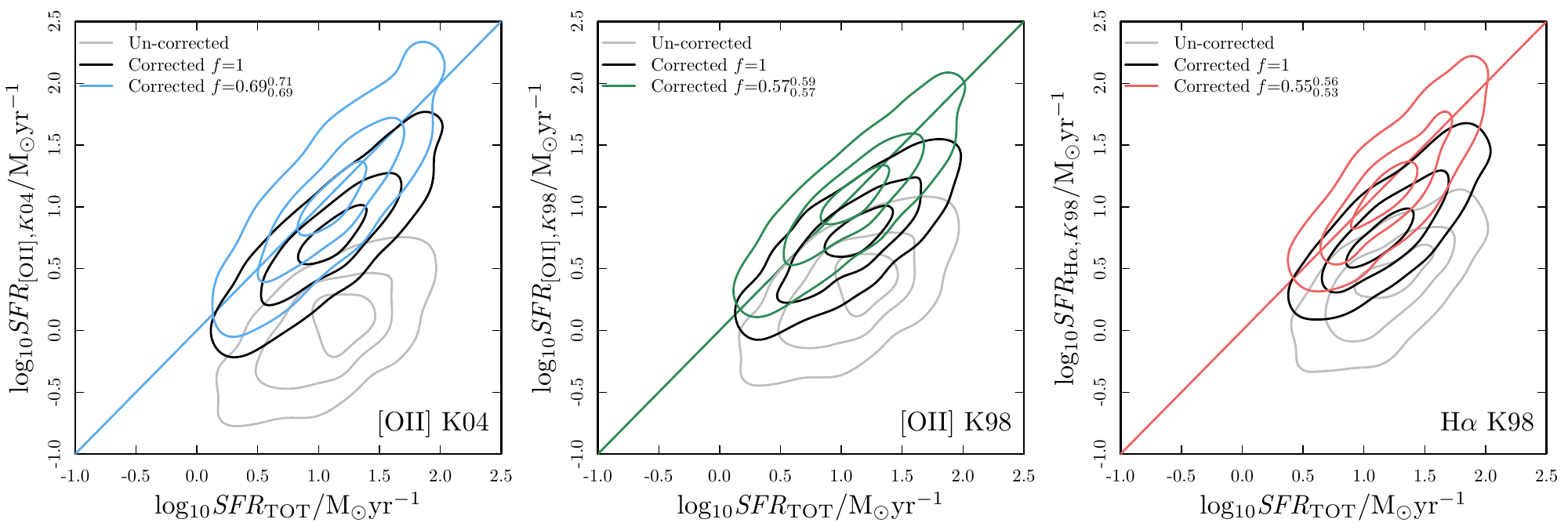}
\caption{From left to right identification of the $f$-factors for \OII, using the $\mathrm{\it SFR}_{\OII}$ calibrations by \citetalias{2004AJ....127.2002K} and \citetalias{1998ARA&A..36..189K}, and for the \hal, respectively. In each case, we display the  comparison between  (SFR$_{\mathrm{TOT}}$ and the {\it SFR} obtained from each EL without correction for dust (gray), the {\it SFR} corrected using the \citetalias{2000ApJ...533..682C} attenuation curve and no $f$-factor ($f$-factor$=$1; black), and the {\it SFR} corrected using the same attenuation curve, and the $f$-factor that minimizes the residuals (blue, green,  and red respectively for {\it SFR}$_{\protect{\OII},K04}$, {\it SFR}$_{\OII,K98}$, and {\it SFR}$_{H\alpha,K98}$). In the three panels, the contours display the percentiles P16$^{th}$, P50$^{th}$, and P84$^{th}$.
}\label{fig:sfr_sfr_corr_ir}
\end{figure*}

\subsection{The dust attenuation correction: the framework} \label{sec:attenuation}

The impact of dust on the light intrinsically emitted by a 
galaxy at a given wavelength ($\lambda$) is usually presented as
\begin{equation}
L_{\mathrm{obs}}(\lambda) = L_{\mathrm{int}}(\lambda) \times 10^{-0.4 A_{\lambda}}, \label{eq:ll_attenuation}
\end{equation}
\noindent where $L_{\mathrm{obs}}(\lambda)$ and 
$L_{\mathrm{int}}(\lambda)$ are the dust-obscured and the 
intrinsic luminosities, and $A_{\lambda}$ is 
the dust attenuation. 

In turn, the attenuation can be parametrized as 
\begin{equation}
A_{\lambda} = E(B-V) \times k(\lambda) \label{eq:attenuation}
\end{equation}
\noindent where $E(B-V)$ is the color excess or reddening [$E(B-V)\equiv A_{B}-A_{V}$, where $A_{B}$ and $A_{V}$ are the attenuations in the $B$ and $V$ bands] and $k(\lambda)$ is the so-called total formulation of the attenuation curve (e.g., see the review by \citealt{2001PASP..113.1449C}). 

Frequently, the attenuation is given as a function of the value of the total attenuation curve in the $V$ band, known as $R_{V}$ ($R_V\equiv k_V=A_{V}/E(B-V)$): 
\begin{equation}
A_{\lambda} = A_V \times k(\lambda)/R_V \label{eq:attenuation2}
\end{equation}
$R_V$ is used to effectively parametrize the observed variations in the attenuation curves of galaxies  (e.g., \citealt{2018ApJ...859...11S}). For the well studied Milky Way (MW) and Large Magellanic Cloud (LMC) attenuation curves it is customary to use an average $R_V=3.1$, although different lines of sight through the diffuse ISM give values ranging from 2 to 6 \citep{1989ApJ...345..245C}. The curve by \citetalias{2000ApJ...533..682C} is usually associated to $R_V=4.05$, which is the average value found for a sample of low redshift starburst galaxies with a galaxy-to-galaxy scatter of $\Delta R_V$$=$0.8. 

{A differential attenuation between stellar and nebular emissions implies a difference in the dust curve that affects these components or/and a differential reddening. This latter is frequently quantified by the so-called $f$-factor (\citetalias{2000ApJ...533..682C}), which is defined as the ratio between the reddening of stellar continuum  [$E(B-V)_{\mathrm{star}}$] and nebular [$E(B-V)_{\mathrm{neb}}$] emission:
\begin{equation}
E(B-V)_{star} =\,\, f \times E(B-V)_{neb}. \label{eq:differential}
\end{equation}
}
\noindent{The differential reddening has been quantified in several works. The most widely used value of $f$ is 0.44$\pm$0.03, which was obtained by \citetalias{2000ApJ...533..682C} for the local universe.}

\begin{table}
	\centering
	\caption{We report the $f$-factors obtained for the different subsamples described in Table~\ref{tab:samples} for \OII\,following the calibration by \citetalias{2004AJ....127.2002K} and \citetalias{1998ARA&A..36..189K}, and the \hal\,through the calibration by \citetalias{1998ARA&A..36..189K}. The results are given in the shape of the median and the percentiles P16$^{th}$ and P84$^{th}$. We consider the \citetalias{2000ApJ...533..682C} attenuation curve for both stellar continuum and nebular emission.}
	\label{tab:f}
	\begin{tabular}{l|ccc} 
		\hline
		Sample of & 
		$f_{\mathrm{[OII]K04}}$ & $f_{\mathrm{[OII]K98}}$ & $f_{\mathrm{H}\alpha}$\\
		\OII-emitters &  &  &   \\
		(1) & (2) & (3) & (4) \\
		\hline
		 IR-detected  &  0.69 $^{ 0.71 }_{ 0.69 }$ & 0.57 $^{ 0.59 }_{ 0.57 }$ & \\
		 IR-detected \& H$\alpha$ & 0.69 $^{ 0.69 }_{ 0.67 }$ & 0.56 $^{ 0.57 }_{ 0.55 }$ &  {0.55$^{ 0.56 }_{0.53 }$} \\
		 \hline
	\end{tabular}
\end{table}

Analogously to Equation~\ref{eq:ll_attenuation}, the attenuation correction for any SF tracer can be derived by comparing the obscured {\it SFR}$^{\mathrm{obs}}_{\mathrm{tracer}}$ with the reference or total one. Thus, 
\begin{equation}
SFR_{\mathrm{TOT}}  = {\it SFR}^{\mathrm{obs}}_{\mathrm{tracer}} \times 10^{0.4*A_{\mathrm{tracer}}} \label{eq:tracer}
\end{equation}
where $A_{\mathrm{tracer}}$ is the attenuation of the luminosity of the SF tracer that we want to correct for dust attenuation. 

When the tracer is the UV (e.g., monochromatic luminosity at 1600\AA), we can express the UV attenuation as follows: 
\begin{equation}
A_{\mathrm{UV}} = 2.5 \times \mathrm{log}_{10} \left( \frac{SFR_{\mathrm{TOT}}}{SFR_{\mathrm{UV}}} \right)\label{eq:A}
\end{equation}
This quantity is sometimes referred to as $A_{\mathrm{IRX}}$ as it can be also expressed as a function of the infrared excess ({\it IRX}, \citealt{1999ApJ...521...64M}). In the case of the {\it SFR}s traced by ELs, we can express their attenuation as 
\begin{equation}
A_{\mathrm{EL}} = 2.5 \times \mathrm{log}_{10} \left( \frac{SFR_{\mathrm{TOT}}}{SFR_{\mathrm{EL}}} \right)\label{eq:Ael}
\end{equation}
{The data set we have in hand allows us to pivot on $A_{\mathrm{UV}}$ to derive the attenuation needed to correct any {\it SFR} traced by EL luminosities.} Using Equations~\ref{eq:attenuation} and \ref{eq:differential}, and attenuation/extinction curves, we can express the attenuation of these lines as 
\begin{equation}
A_{\mathrm{EL}} = A_{\mathrm{UV}} \frac{k(\lambda_{\mathrm{EL}})}{k(\lambda_{\mathrm{UV}})} \times  \frac{1}{f}\label{eq:AEL}
\end{equation}

\noindent We assume the \citetalias{2000ApJ...533..682C} attenuation curve with $R_V$=4.05 for both stellar continuum and nebular emission.
We note that the wavelength at which it is correct to evaluate the attenuation curve in Equation~\ref{eq:AEL} depends on the recipe used to convert luminosities into {\it SFR}s. For instance, in the case of the \citetalias{1998ARA&A..36..189K} calibration for the $\mathrm{\it SFR}_{\OII}$, the right wavelength on which calculate the $k$($\lambda_{\mathrm{EL}}$) is the wavelength of the \hal\,line. {This is because this recipe relies on that of the $\mathrm{\it SFR}_{\mathrm{H}\alpha}$, where the luminosity of \hal\, is changed with the luminosity of \OII\, divided by an average value of the \OII/H$\alpha$ ratio, with \OII\,and \hal\,fluxes not being dust-corrected (see  \citetalias{1998ARA&A..36..189K} and  \citetalias{2004AJ....127.2002K}). In the case of the  calibration by  \citetalias{2004AJ....127.2002K}, the wavelength at which the attenuation curve is evaluated is the actual wavelength of the \OII, due to the fact that the calibration is built adopting an average dust-corrected \OII/H$\alpha$ ratio.}

\subsection{The \lowercase{{\it f}}-factor for \OII\, and \hal}\label{subsec:ffactor}
In Figure~\ref{fig:sfr_sfr_corr_ir}, we report the $f$-factor found for the \OII\, ($f_{\OII,\mathrm{K04}}$ and $f_{\OII,\mathrm{K98}}$) and \hal\,ELs ($f_{\mathrm{H}\alpha,\mathrm{K98}}$) considering the three {\it SFR} calibrations in Section~\ref{seq:SFREL}. 
 In practice, we minimize the residuals between $SFR_{\mathrm{TOT}}$ and the dust-corrected  {\it SFR} probed by \OII\, and \hal\,($SFR^{\mathrm{corr}}_{\OII, K04}$, $SFR^{\mathrm{corr}}_{\OII, K98}$ and $SFR^{\mathrm{corr}}_{\mathrm{H}\alpha,K98}$), assuming the \citetalias{2000ApJ...533..682C} attenuation curve with $R_V$\,$=$\,4.05. We use Monte Carlo simulations to assess the uncertainties of the results. This means that we perform 1000 iterations of the minimization considering in each repetition a sample of simulated values of {\it SFR} that we generate randomly within Gaussian probability distributions centered in each original {\it SFR} data-point and a $\sigma$ corresponding to their uncertainty. The results are given in the shape of the median and percentiles (P16$^{th}$ and P86$^{th}$) of the output of the 1000 iterations (see Table~\ref{tab:f}). 

 Figure~\ref{fig:sfr_sfr_corr_ir} displays the need for an $f$-factor for both \OII\, and \hal\,ELs to describe {the reddening of ELs when compared to that of the stellar continuum}. The $f$-factors obtained are 0.69$^{0.71}_{0.69}$ and 0.57$^{0.59}_{0.57}$ for the \OII\, (\citetalias{2004AJ....127.2002K} and \citetalias{1998ARA&A..36..189K} calibrations, respectively). When using only the \OII-emitters with an \hal\,detection, the values of the $f$-factor for the \OII\, are 0.69$^{0.69}_{0.67}$ and 0.56$^{0.57}_{0.55}$ (\citetalias{2004AJ....127.2002K} and \citetalias{1998ARA&A..36..189K} calibrations, respectively). We find an $f_{\mathrm{H\alpha, K98}}$ equal to {0.55$^{0.56}_{0.53}$}. The difference between $f_{\mathrm{H}\alpha}$ (and $f_{\mathrm{\OII, K98}}$) and $f_{\mathrm{\OII, K04}}$ appears to be significant. It is worth noting that the uncertainties of these values are not informative of the range over which the $f$-factors calculated for individual galaxies expand. 

Our results for \hal\, and \OII\,are consistent with the $f$-value found by \citetalias{2000ApJ...533..682C}. {The canonical value given by the original \citetalias{2000ApJ...533..682C} work is $f$$=$0.44. However, this corresponds to $f$$=$0.58 if the
\citetalias{2000ApJ...533..682C} law were assumed for both nebular and stellar continuum emission  (\citealt{2014ApJ...795..165S},  \citealt{2015ApJ...807..141P}).}
They are also comparable with the findings published by \citet{2014ApJ...788...86P}, where an $f$$=$0.54$^{0.78}_{0.44}$ for a sample of 163 star forming galaxies observed by the 3D-HST survey between 1.36$<$$z$$<$1.5 by relying on Balmer decrement measurements. Also, \citet{2015A&A...582A..80T} found $f$$=$ 0.50 for a sample of $\sim$200 IR-detected [OII]-emitters at 1$<$$z$$<$1.3. 

{It is worth noting that the value of the differential reddening can vary noticeably if different dust curves are assumed for either or both the stellar continuum and the nebular emission. For instance, if we assume a \citetalias{2000ApJ...533..682C} curve for the former and a \citet[][]{1999PASP..111...63F} curve with R$_v$=3.10 for the latter, we obtain $f_{\mathrm{\OII, K04}}$$=$0.73$^{0.75}_{0.72}$ and 
$f_{\mathrm{H}\alpha}$$=$0.51$^{0.53}_{0.50}$. Instead, assuming a \citet[][]{1989ApJ...345..245C} curve for the nebular emission yields $f_{\mathrm{\OII, K04}}$$=$0.57$^{0.58}_{0.56}$ and 
$f_{\mathrm{H}\alpha}$$=$0.41$^{0.43}_{0.40}$. If we now consider a steeper dust curve for the stellar continuum, such as the typical extinction curve in the Small Magellanic Cloud published by \citet{1984A&A...132..389P}, and assume the \citet[][]{1989ApJ...345..245C} extinction curve for the emission lines, we obtain 
$f_{\mathrm{\OII, K04}}$$=$0.49$^{0.48}_{0.50}$ and 
$f_{\mathrm{H}\alpha}$$=$0.39$^{0.40}_{0.38}$. In this later case, the assumption of the \citetalias{2000ApJ...533..682C} curve for the nebular emission gives $f_{\mathrm{\OII, K04}}$$=$0.59$^{0.61}_{0.59}$ and 
$f_{\mathrm{H}\alpha}$$=$0.47$^{0.48}_{0.46}$. }

\section{Dependence of the \lowercase{$f$}-factor on redshift and physical parameters}
\label{sec:ffactor_dep}

\begin{figure*}
\includegraphics[width=\linewidth]{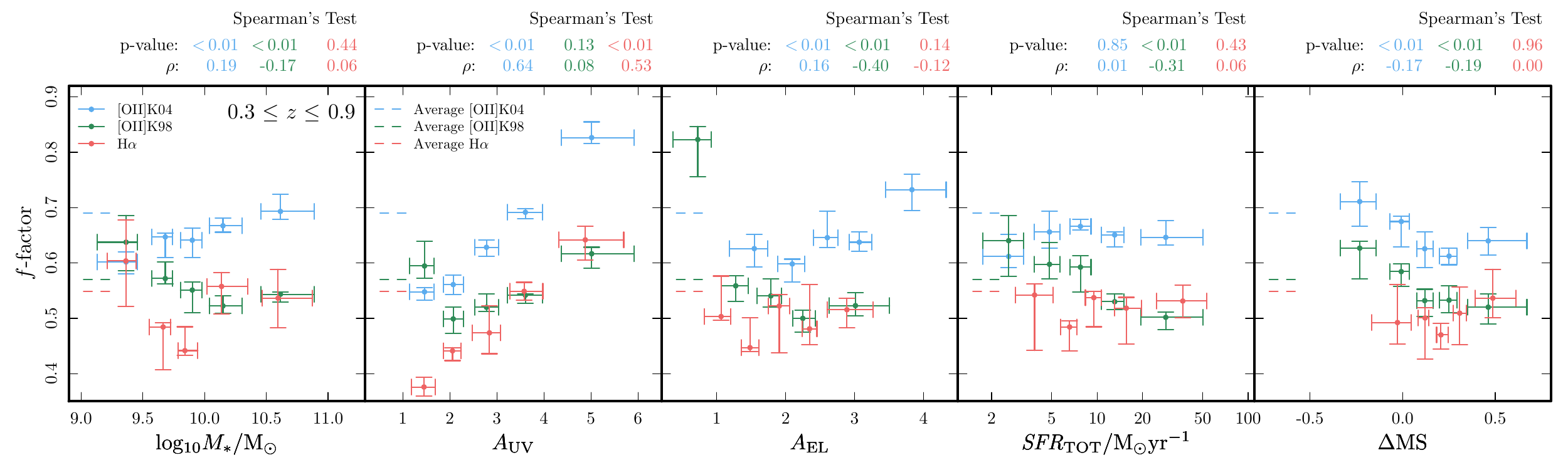}
\includegraphics[width=\linewidth]{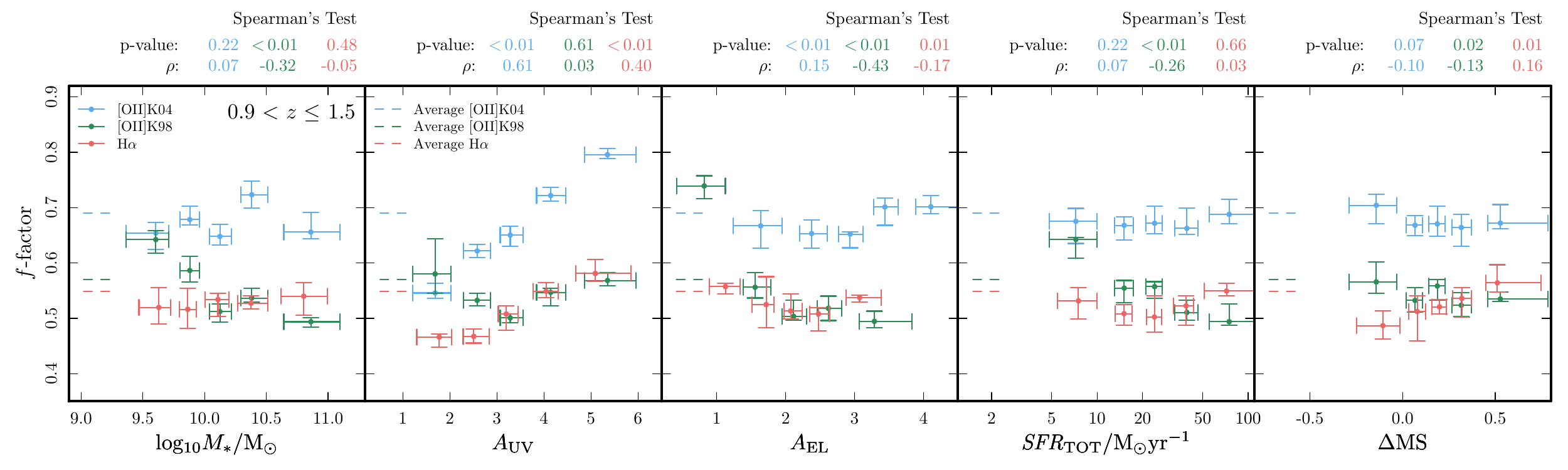}
\caption{{Dependence of the $f$-factor on different physical properties in two redshift bins. From left to right the panels display the dependence on stellar mass, UV attenuation, EL attenuation, total {\it SFR}, and the distance to the main sequence. 
Upper and bottom panels show the results for two redshift bins. The values of $f$ are obtained for each subsample by minimizing the residuals between the {\it SFR}$_{\mathrm{TOT}}$ and the {\it SFR}$^{\mathrm{corr}}_{\OII,K04}$ (blue data points), {\it SFR}$^{\mathrm{corr}}_{\OII,K98}$ (green data points), {\it SFR}$^{\mathrm{corr}}_{H\alpha,K98}$ (red data points). The error bars are obtained through Monte Carlo simulations. The values obtained for the complete sample (Section~\ref{sec:ffactor}) are marked as horizontal dashed segments. The output of the Spearman's correlation tests performed using the $f$ values of individual galaxies are also displayed over each panel.} }\label{fig:f_dependences}
\end{figure*}

Figure~\ref{fig:f_dependences} shows the average values of the $f$-factor in equally populated bins of different galaxy physical properties {for two redshift bins}. The values per bin are obtained applying the same methodology as in Section~\ref{sec:ffactor}, i.e., minimizing the residuals between the {\it SFR}$_{\mathrm{TOT}}$ and the {\it SFR}$^{\mathrm{corr}}_{\mathrm{EL}}$. We clarify that we do not average the $f$-factors obtained for individual galaxies. {Using this alternative methodology the results do not change significantly}.  We assess the correlations using the Spearman's correlation {test}, which is also displayed in each panel of  Figure~\ref{fig:f_dependences}.

\subsection{Dependence on stellar mass}
{We observe a few interesting features in the panels of Figure~\ref{fig:f_dependences} where the trends with stellar mass are displayed. First of all, we find a significant (p-value $<$0.01) slightly increasing trend of $f_{\mathrm{\OII,K04}}$ with stellar mass in the lowest redshift bin. The correlation appears to vanish at higher redshifts, which could be due to a poorer sampling of smaller stellar masses.} Different papers in literature have found hints of an increasing trend of $f$-factor with stellar mass (e.g. \citealt{2014ApJ...788...86P}, \citealt{2016A&A...586A..83P}). {However, our results are not conclusive considering the shallow trend displayed by our data points. Moreover, we find no significant correlations in the case of H$\alpha$. }
{The Spearman's test returns a high-significance inverse correlation between  $f_{\mathrm{\OII,K98}}$ and the stellar mass throughout the whole redshift range probed. 
The figure shows that the behaviour of $f_{\mathrm{\OII,K98}}$ is rather flat and analogous to that of  $f_{\mathrm{H\alpha,K98}}$ at log$_{10}$($M_{*}$/M$_{\odot}$)$>$10, with decreasing differential reddening (i.e. larger $f$) at log$_{10}$($M_{*}$/M$_{\odot}$)$<$10. 
This inverse correlation can appear puzzling given the results found for $f_{\mathrm{\OII,K04}}$. However, it could be explained by an overestimate of {\it SFR}$_{\OII}$ that affects low-mass galaxies when the \citetalias{1998ARA&A..36..189K} calibration is applied. Low-mass galaxies can display (dust un-corrected) \OII/\hal\,ratios 2 times larger than the ratio assumed by the calibration (see Figure~3 in \citetalias{2004AJ....127.2002K}).  If {\it SFR}$_{\OII}$ is overestimated, then, the dust correction needed to recover the estimate of the total star formation activity is smaller, which implies reduced differential reddening, i.e., values of $f$-factor closer to 1.} 

{As it was mentioned in Section~\ref{seq:SFREL}, the calibrations of {\it SFR}$_{\OII}$ that we use rely on an assumed average \OII/\hal\,ratio. However, the value of this ratio is dependent on different physical properties of galaxies and can change considerably. In particular, \citetalias{2004AJ....127.2002K} find that attenuation and metallicity effects can lead to significant disagreements between {\it SFR}$_{\OII}$ and {\it SFR}$_{\mathrm{H}\alpha}$.  The  \citetalias{2004AJ....127.2002K} calibration that we consider does not contain a reddening assumption, however, it implicitly assumes an average excitation state of the gas and an average metallicity. The tight mass–metallicity relation (e.g., \citealt{2004ApJ...613..898T}, \citealt{2013ApJ...771L..19Z}) observed in samples of galaxies throughout cosmic times implies that the average  \OII/H$\alpha$ ratio varies across the range of stellar mass we probe. In 
general, smaller values of \OII/H$\alpha$ are expected for high metallicities (i.e. more massive galaxies), although in the low metallicity regime this EL ratio can become less sensitive to variations of oxygen abundance and also decrease with decreasing oxygen abundance, depending on the calibration considered (see \citetalias{2004AJ....127.2002K}, and references therein). 
If the average dust-corrected \OII/\hal\,ratio obtained by \citetalias{2004AJ....127.2002K} were larger than the average ratio for a particular sample of galaxies, the recipe would underestimate their average values of {\it SFR}$_{\mathrm{\OII,K04}}$. In such case, the resulting $f_{\mathrm{\OII,K04}}$ would be smaller in order to counter the difference between {\it SFR}$_{\mathrm{TOT}}$ and {\it SFR}$_{\mathrm{\OII,K04}}$. Therefore, the mass–metallicity relation would likely translate into a decreasing trend of $f_{\mathrm{\OII,K04}}$ with stellar mass.}

\subsection{Dependence on attenuation}
{We find significant positive correlations between $f_{\mathrm{\OII,K04}}$ and $f_{\mathrm{H\alpha,K98}}$ and the UV attenuation in both redshift bins. 
The behaviour of $f_{\mathrm{\OII,K98}}$ resembles that of $f_{\mathrm{H\alpha,K98}}$ for attenuations larger than $A_{\mathrm{UV}}$$=$3. At smaller attenuations, the trend flattens and seems to  undergo an upturn. This not-monotonic behaviour of the relation between $f_{\mathrm{\OII,K98}}$ and $A_{\mathrm{UV}}$ is consistent with the results found for the stellar mass, and  it translates into low-significance correlation results for the Spearman's Test.}  
{We find overall weaker correlations between $f_{\mathrm{\OII,K04}}$ and $f_{\mathrm{H\alpha,K98}}$ and the attenuation of these emission lines in both redshift bins. }

The correlation between $f$ and $A_{\mathrm{UV}}$ could explain the smaller differential reddening that studies based on FIR detected galaxies (i.e., dustier on average) preferentially obtain (e.g., \citealt{2016A&A...586A..83P}). We note that the selection of \OII-emitters biases the sample towards less obscured systems, otherwise, the EL would be too faint for a detection. For instance,  \citet{2013MNRAS.430.1042H} used a dual narrow-band survey strategy to select 809 SFGs at $z$$=$1.47 with \hal\,and \OII\, emission, and found that \OII-selected narrow-band emitters are typically dust-poorer systems than \hal-selected ones. They also found that this bias increases with redshift.

{The total dust column density along the line of sight appears to be tightly liked to the attenuation curve slope. Larger (smaller) optical depths correspond to grayer (steeper) attenuation curves (e.g., \citealt{2018ApJ...859...11S}, \citealt{2018ApJ...869...70N}). Variations in the FUV slope of the attenuation curves in this direction could imply an enhanced dust absorption of ionizing Lyman continuum photons for galaxies less obscured, which would translate into an  underestimation of the {\it SFR} as traced by \hal\,and \OII\,(e.g., \citealt{2016A&A...586A..83P}). Given our methodology to calculate the differential reddening, this effect would eventually lead to smaller values of $f$ for less obscured galaxies, which is consistent with the trend we observe in Figure~\ref{fig:f_dependences}.}

\subsection{Dependence on {\it SFR}}
{We do not find significance correlations between $f_{\mathrm{\OII,K04}}$ and $f_{\mathrm{H\alpha,K98}}$ and the SFR$_{\mathrm{TOT}}$. The significant (yet weak) correlation between $f_{\mathrm{\OII,K04}}$ and stellar mass at low redshifts does not translate into a significant dependence on {\it SFR}.}
 Contrarily, $f_{\mathrm{\OII,K98}}$ displays a significant negative correlation with the total SF activity, which is probably linked to the peculiar behaviour of $f_{\mathrm{\OII,K98}}$ for low-mass and low-attenuation systems.

\subsection{Dependence on the distance to the main sequence}
{The right-hand panel in Figure~\ref{fig:f_dependences} shows the dependence of the $f$-factor with the distance to the MS by \citetalias{2019ApJS..243...22B}. At low redshifts we find significant inverse correlations between $f_{\OII,K04}$ and $f_{\OII,K98}$ and $\Delta$MS. No correlation is found between $f_{\mathrm{H\alpha,K98}}$ and $\Delta$MS. In the highest redshift bin, however, we find that only  $f_{\mathrm{H\alpha,K98}}$ seems to moderately correlate with with $\Delta$MS (p-value $=$0.01 and $\rho$$=$0.16).}   
We point out the fact that a sample selection based on SF tracers represents a horizontal cut in the MS plane. This introduces a stellar mass bias in the $\Delta$MS bins, with negative (positive) $\Delta$MS bins being populated by more (less) massive galaxies. If this dependence in mass were the main or only parameter playing a role in this plot, we would expect a monotonically decreasing trend of $f$ with $\Delta$MS. Some works have identified a decrease in the amount of extra attenuation suffered by ELs with increasing  specific {\it SFR}  ({\it sSFR}$=${\it SFR}/$M_{*}$; e.g., \citealt{2011MNRAS.417.1760W}, \citealt{2014ApJ...788...86P}). This would imply an increasing trend of $f$ with $\Delta$MS. 

\subsection{Dependence on redshift}
{The $f$-factor does not appear to significantly correlate with redshift for any of the ELs. We arrive at this conclusion by comparing the upper and lower panels in Figure~\ref{fig:f_dependences}.}  
Although there is not consensus in literature, there is a growing evidence showing that at $z$$>$1, the discrepancy between the nebular and stellar attenuation could be smaller (i.e., approaching unity $f$ at higher redshifts; \citealt{2013ApJ...777L...8K}, \citealt{2016ApJ...828..107R}, \citealt{2016A&A...586A..83P}, \citealt{2019ApJ...871..128T}, \citealt{2020A&A...635A.119C}). However, in some cases, the trend is based on the comparison with the widely used $f$-factor value 0.44 found by \citetalias{2000ApJ...533..682C} for a sample of local starbursts. Some of the redshift trends could be the result of observational biases in high-redshift samples which normally are populated by more obscured and actively star-forming galaxies (e.g., \citealt{2011MNRAS.417.1760W}, \citealt{2014ApJ...788...86P}, \citealt{2017MNRAS.472.4878V}). 

Increasing trends with redshift are also often used to make predictions of ELG observability in spectroscopic and spectro-photometric surveys (e.g., \citealt{2019A&A...631A..82I}). For instance, in their recent work, \citet{2020MNRAS.494..199S} propose a redshift evolution of $f$ with the law $f$$=(0.44\pm0.2) \times z$ (at $z$$<$2.8). Despite the fact that we do not find a redshift dependence for $f$, we find values that agree overall with their prediction at the same redshifts. That been said, their $f$ values seem to be too low for \OII\,at the lowest redshifts and too high for \hal\,at the largest redshifts. This is in agreement with the misbehaviour of the predictions pointed out by the authors: their \OII\,line is slightly over-corrected at low-$z$. 

\section{Discussion}\label{sec:discuss}

\begin{figure*}
\includegraphics[page=1,width=\linewidth]{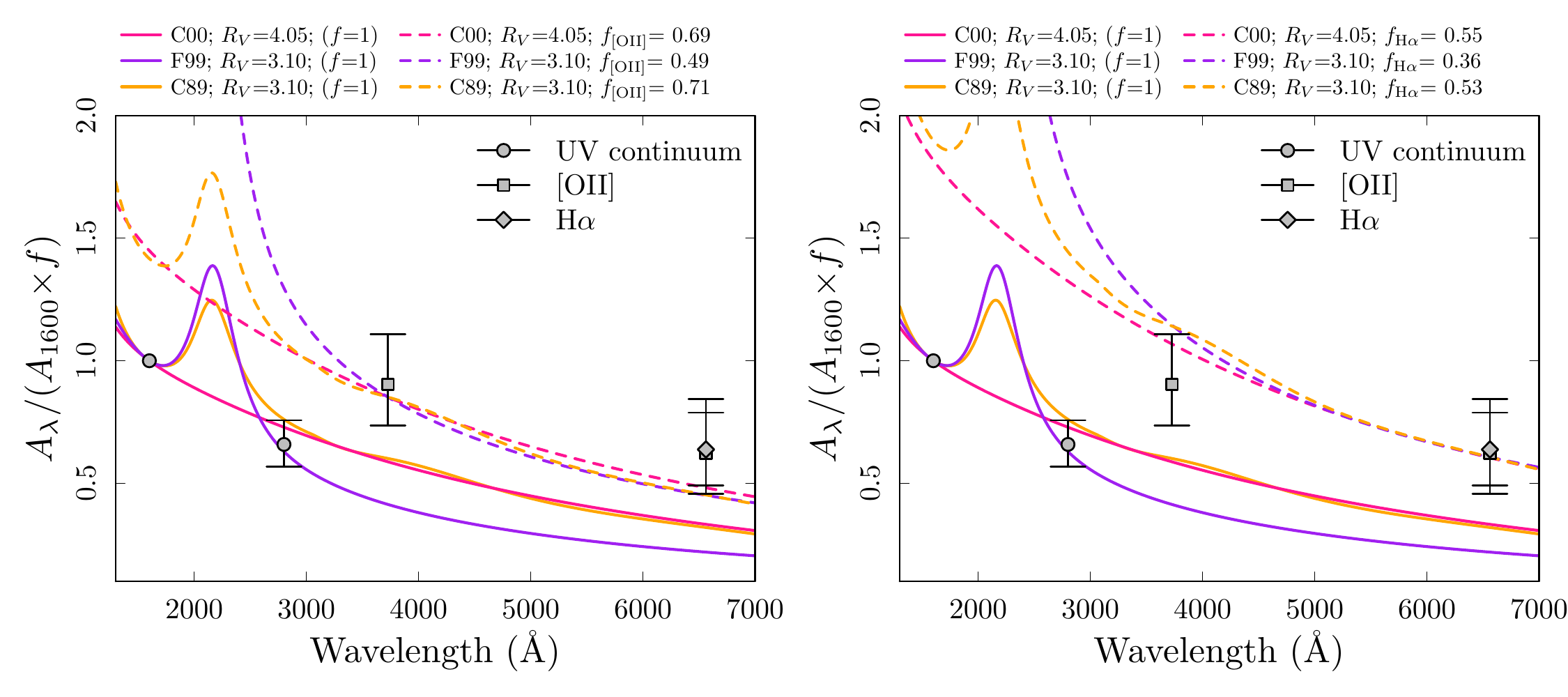}
\caption{Comparison between some of the attenuation/extinction curves more frequently used in the literature normalized to their value at 1600\AA, and the average observed data points at the wavelengths of the {\it SFR} tracers analysed (i.e., UV continuum at 1600\AA\,and 2800\AA, \OII\,, and \hal). The attenuation curves shown  (continuous lines) are the following: \citet[][C00]{2000ApJ...533..682C} with R$_v$=4.05 (starbursts); \citet[][C89]{1989ApJ...345..245C} with R$_v$=3.10 (MW); \citet[][F99]{1999PASP..111...63F} with R$_v$=3.10. 
{The data points (gray symbols) represent the median values (P16$^{th}$ and P84$^{th}$ are displayed in the shape of error bars) for the sample of 396 IR-detected \OII-emitters with log$_{10}$($M_{*}/M{\odot}$)$>$9 and an available H$\alpha$ measurement.} We note that the data point of the attenuation derived for the \OII\,using the calibration by \citetalias{1998ARA&A..36..189K} is placed at the wavelength of the \hal\,for consistency. The dashed lines represent the same attenuation curves elevated by a factor 1/$f_{\OII}$ (left-hand panel) and 1/$f_{\mathrm{H}\alpha}$ (right-hand panel). 
}\label{fig:A_l_A_EL}
\end{figure*}

The first key result of our study is that ELs in intermediate-to-high redshift IR-detected ELGs appear to suffer larger amounts of attenuation than the stellar continuum. This result has been found in several studies of galaxies at low and high redshifts (e.g., \citetalias{2000ApJ...533..682C}, \citealt{2014ApJ...788...86P}, \citealt{2015ApJ...806..259R},
\citealt{2015A&A...582A..80T},
\citealt{2020ApJ...902..123R}, \citealt{2020ApJ...899..117S}, \citealt{2013ApJ...771...62K}).
The need of a larger dust attenuation correction for the nebular emission could be explained by the following two-component dust model. All stars experience a modest attenuation due to the diffuse ISM dust. Additionally the population of young and massive stars together with the nebular emission they trigger are embedded in dense and dusty molecular clouds.
Some authors have found a strong correlation between the magnitude of this differential attenuation and the {\it sSFR} (e.g., \citealt{2011MNRAS.417.1760W}, \citealt{2014ApJ...788...86P}). The way in which this correlation is explained is the following (see Figure 5 in  \citealt{2014ApJ...788...86P}). In galaxies with the highest {\it sSFR}s, the continuum is dominated by young and massive stars, presumably affected by the dust in the birth clouds in a way similar to the ELs. In the case of the lowest {\it sSFR}s, the ELs and the continuum features would be attenuated by dust with very different properties, leading to larger differential attenuations. However, our results do not follow this behaviour. They rather give evidence for a weak dependence of the $f$-factor on the stellar mass, the star formation activity and the {\it sSFR} while they point towards a {stronger dependence on the UV attenuation.}. 

\subsection{Nebular emission attenuation curve}\label{sec:nebular_Acurve}

In this section, we explore possible average {dust} curves that could modulate the nebular emission of our sample of galaxies. Figure~\ref{fig:A_l_A_EL} displays the attenuation curve by \citetalias{2000ApJ...533..682C} normalized to their value at 1600~\AA. We also include the extinction curves by \citet[][]{1989ApJ...345..245C} with R$_v$=3.10 and \citet[][]{1999PASP..111...63F} with R$_v$=3.10. 

 We use Equations~\ref{eq:A} and~\ref{eq:Ael}  to derive observed attenuations for the UV continuum and the ELs. The figure includes the average (P50$^{th}$, P16$^{th}$, and P84$^{th}$) values of these attenuations once they are normalized to the former. We only include the sample of IR-detected \OII-emitters with a detection in \hal\, so that all the data points in Figure~\ref{fig:A_l_A_EL} are obtained with the same sample of galaxies. Note that the data point obtained using the {\it SFR}$_{\mathrm{\OII,K98}}$ is placed at the wavelength of \hal. 
 The left-hand and right-hand panels
 also give the attenuation/extinction curves once they are scaled by the factors that make them fit  A$_{\OII}$/A$_{\mathrm{UV}}$ and A$_{H\alpha}$/A$_{\mathrm{UV}}$, respectively. 
 
Figure~\ref{fig:A_l_A_EL} shows that $f$-factors are needed to correctly evaluate the attenuation of ELs. The figure also shows that the data points of \OII\,and \hal\, are overall compatible with the extinction curves by \citet[][]{1989ApJ...345..245C} and \citet[][]{1999PASP..111...63F}, and the \citetalias{2000ApJ...533..682C}  attenuation curve broadly used in the literature, also at high redshifts. This is in agreement with the recent work by \citet{2020ApJ...902..123R}, in which they use the first five low-order Balmer ELs measured in the composite spectra of 532 galaxies at 1.4$<$$z$$<$2.6 observed by the MOSFIRE Deep Evolution Field survey. Our results suggest that the integrated dust absorption and scattering properties in our sample do not depart significantly from those of the Milky Way or low-redshift starbursts. Finally, we note that this result gives hints on the average behaviour of the whole sample of IR-detected \OII-emitters. However, galaxy-to-galaxy differences in the shape of the {dust} curve have been reported in the literature (e.g., \citealt{2018MNRAS.475.2363T}, \citealt{2018ApJ...859...11S}).

\subsection{Caveats: the many factors affecting the {\textit f}-factor}\label{sec:caveats}
Differences in attenuation between nebular and stellar continuum emission components may well exist in nature, because of departures from co-spatiality, as widely entertained in the literature, that led to the introduction of the $f$-factor (\citetalias{2000ApJ...533..682C}). Numerous attempts to probe such difference have been carried out. The optimal way to approach this task is to compare two different direct measures of attenuation, such as the Balmer decrement for the lines (e.g., \citealt{2020ApJ...902..123R}) and the UV attenuation as derived from the UV slope. In our work, we calculate $f$ by enforcing equality between two {\it SFR} for the same galaxy, as derived from two different {\it SFR} diagnostics. This methodology relies on the fact that the calibrations of the {\it SFR} for the different tracers give the same {\it SFR} value once they are corrected {for} attenuation. Instead, this assumption can be challenged under certain circumstances. Even in the aforementioned ideal approach, it is not easy to disentangle the many effects subsumed in the resulting $f$-factor. Among them, we highlight: 
\begin{itemize}
\item The mentioned lack of co-spatiality between line and UV continuum emitting regions.
\item The actual attenuation curve could be different from the adopted one. The $f$-factors derived here rely on the assumption that the attenuation curve is the \citetalias{2000ApJ...533..682C} one and the same for all galaxies, whereas there are indications that it may depend somewhat on the physical properties of galaxies (e.g., \citealt{2018ApJ...859...11S}, \citealt{2021ApJ...909..213K}).
\item Possible systematic errors in the coefficients linking the luminosity to the {\it SFR}. For instance, in the case of \OII\, as a {\it SFR} indicator, differences in oxygen abundance.  Assumptions included in {\it SFR} calibrations are numerous and intricate (e.g., \citealt{2004AJ....127.2002K}). {In particular, effects such as escape and absorption of ionizing photons are subsumed in the empirical calibration of the {\it SFR}-line luminosity relation, but such calibration strictly apply to the specific galaxies used in the calibration and may not apply to the whole population of galaxies, as assumed.}
{Bursty SF events can also lead to discrepancies in estimates of {\it SFR} based on indicators that probe SF on different timescales.}
\item An offset between the bulk of SF which is extremely attenuated, and the UV and line emitting regions which happens to be in the least attenuated lines of sight (e.g., \citealt{2017ApJ...838L..18P}). Some authors argue that this issue could be linked with the difference in size and counterpart offsets displayed by the submillimetric/radio emission and the \hal\,maps of strongly star-forming galaxies (e.g. \citealt{2009A&A...505.1017G}, \citealt{2018ApJ...867...92S},   \citealt{2020A&A...635A.119C}). Also, some authors have found evidences for an underestimate of the $A_{\mathrm{H}\alpha}$ due to the enhanced optical thickness of the line, that could be explained by high dust column densities within HII regions (e.g., \citealt{2013A&A...553A..85P}).
\item The line flux depends on attenuation because of two entirely different physical reasons: (a) individual \OII/\hal\,photons are
absorbed/scattered, and (b) UV ionizing (Lyman continuum) photons are absorbed by dust before ionizing hydrogen and oxygen (\citealt{2016A&A...586A..83P}). Therefore, the line flux depends also on the absorption in the Lyman continuum, where it is maximum and where different attenuation/extinction curves diverge. This situation depends on the distribution of the dust within the star-forming region (e.g., \citealt{2009ApJ...706.1527B}, \citealt{2016A&A...586A..83P}).  
\end{itemize}

The significant differences between $f$-factor values found in literature for different samples, methodologies, ELs, and considered attenuation curves (e.g., see \citealt{2016A&A...586A..83P} and \citealt{2020ApJ...899..117S} for a summary) is likely due to a combination of all these factors. {Thus, it is difficult to draw a conclusion on what is actually measured by the $f$-factor, or what is the physical origin of the correlations (or their absence) displayed in Figure~\ref{fig:f_dependences}.}

\section{Summary \& Conclusions} \label{sec:conclusions}
We have identified a sample of 706 IR-detected \OII-emitters with log$_{10}$($M_{*}/\mathrm{M}_{\odot}$)$>$9 at 0.3\,$\lesssim$\,$z$\,$\lesssim$\,1.5 in the SHARDS spectro-photometric survey. We have explored the differential attenuation displayed by their \OII\,and \hal\,ELs by comparing the {\it SFR} traced by the nebular emission and a robust independent estimate of their total SF activity that relies on the UV and IR continuum luminosities. The main results and conclusions of our work are the following.
\begin{itemize}
\item An $f$-factor different from 1 is needed in order to properly describe the enhanced attenuation of ELs with respect to stellar continuum in intermediate to high redshift IR-detected ELGs. 
\item We find $f$-factors 0.69$^{0.71}_{0.69}$ and {0.55$^{0.56}_{0.53}$} for \OII\, and \hal, respectively, when considering a \citetalias{2000ApJ...533..682C} attenuation curve with $R_V$$=$4.05 for both stellar continuum and nebular components.
\item The $f$-factor appears to display a significant positive correlation with UV attenuation.  
\item The average impact of dust on  \OII\, and \hal\, appears to be entirely compatible with the \citetalias{2000ApJ...533..682C} nebular attenuation curve. 
\end{itemize}

This work provides information for the correct quantification of the SF activity in ELGs and it is potentially relevant for the success of the present-day and future spectroscopic and spectrophotometric surveys which will unveil large samples of ELGs throughout cosmic times.

\section*{Acknowledgements}
We are grateful to the anonymous referee for her/his constructive comments. L.R.-M. thanks Casiana Muñoz Tuñ\'on for useful suggestions on this manuscript. L.R.-M., G.R., and A.F. acknowledge the support from grant PRIN MIUR 2017-20173ML3WW\_001. 
L.R.-M. also acknowledges funding support from the Universit\`a degli studi di Padova - Dipartimento di Fisica e
Astronomia ``G. Galilei''. P.G.P.-G. and L.C. acknowledges support from Spanish Ministerio de Ciencia, Innovaci\'on y Universidades through grant PGC2018-093499-B-I00. L.C. acknowledges also financial support from Comunidad de Madrid under Atracci\'on de Talento grant 2018-T2/TIC-11612. A.V.G. acknowledges support from the European Research Council through the Advanced Grant MIST (FP7/2017-2020, No 742719). A.P. gratefully acknowledges financial support from STFC through grants ST/T000244/1 and ST/P000541/1. Analyses were performed in R 3.6.1 (\citealt{R}).

\section*{Data Availability} 
The data underlying this article will be shared upon reasonable request to the corresponding author.




\bibliographystyle{mnras}



\appendix

\section{Sample selection}\label{A1}
Figure~\ref{fig:selection} displays a scheme of the technique used to select 
ELGs in each filter and the redshift distribution of the different samples: 
ELGs, \OII-emitters, and among the latter, those selected by spectroscopic 
and photometric redshift.  

\begin{figure*}
\includegraphics[page=1,width=\linewidth]{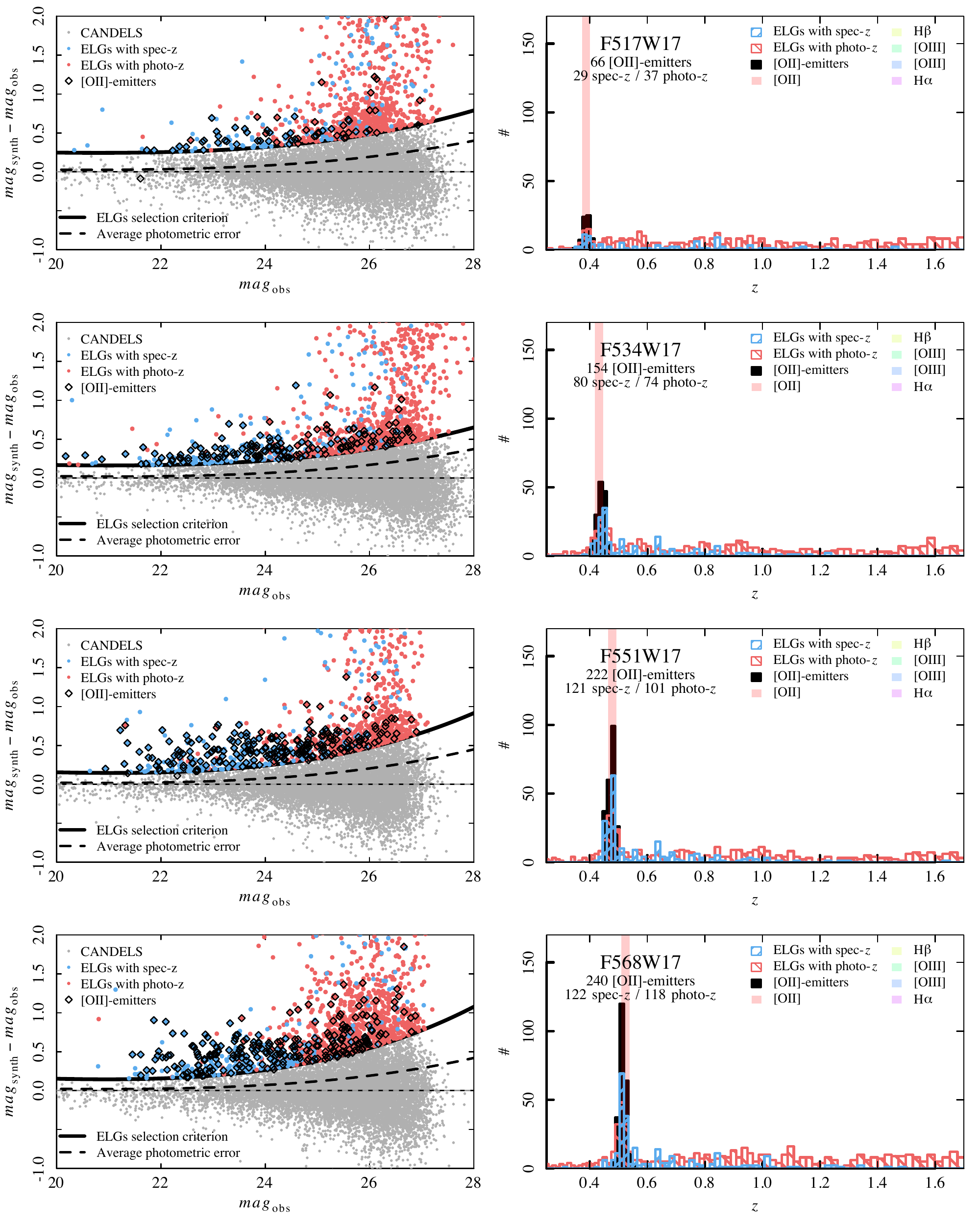}
\caption{{\it Left panels}: Color-magnitude diagram showing ELG candidates 
in each SHARDS filter. Color is defined as the difference between the synthetic
and the observed magnitudes. 
The color threshold which defines the locus of ELGs is marked with a black continuous line. The color equal to zero is marked with a horizontal thin dashed black line. The dashed curve represent the average photometric error. {\it Right panels}: Redshift distribution 
of the ELG candidates throughout the redshift range in which \OII\, falls within the wavelength range over which SHARDS extends. The vertical coloured lines mark the redshift that shifts \OII, H$\beta$, [OIII]4861\AA, [OIII]5007\AA, and \hal\, into the corresponding SHARDS filter. The blue (red) histogram displays the distribution of galaxies with spectroscopic (photometric) redshifts. The black histogram shows the distribution of the \OII-emitters identified. 
}\label{fig:selection}
\end{figure*}
\begin{figure*}
\addtocounter{figure}{-1}
\includegraphics[page=2,width=\linewidth]{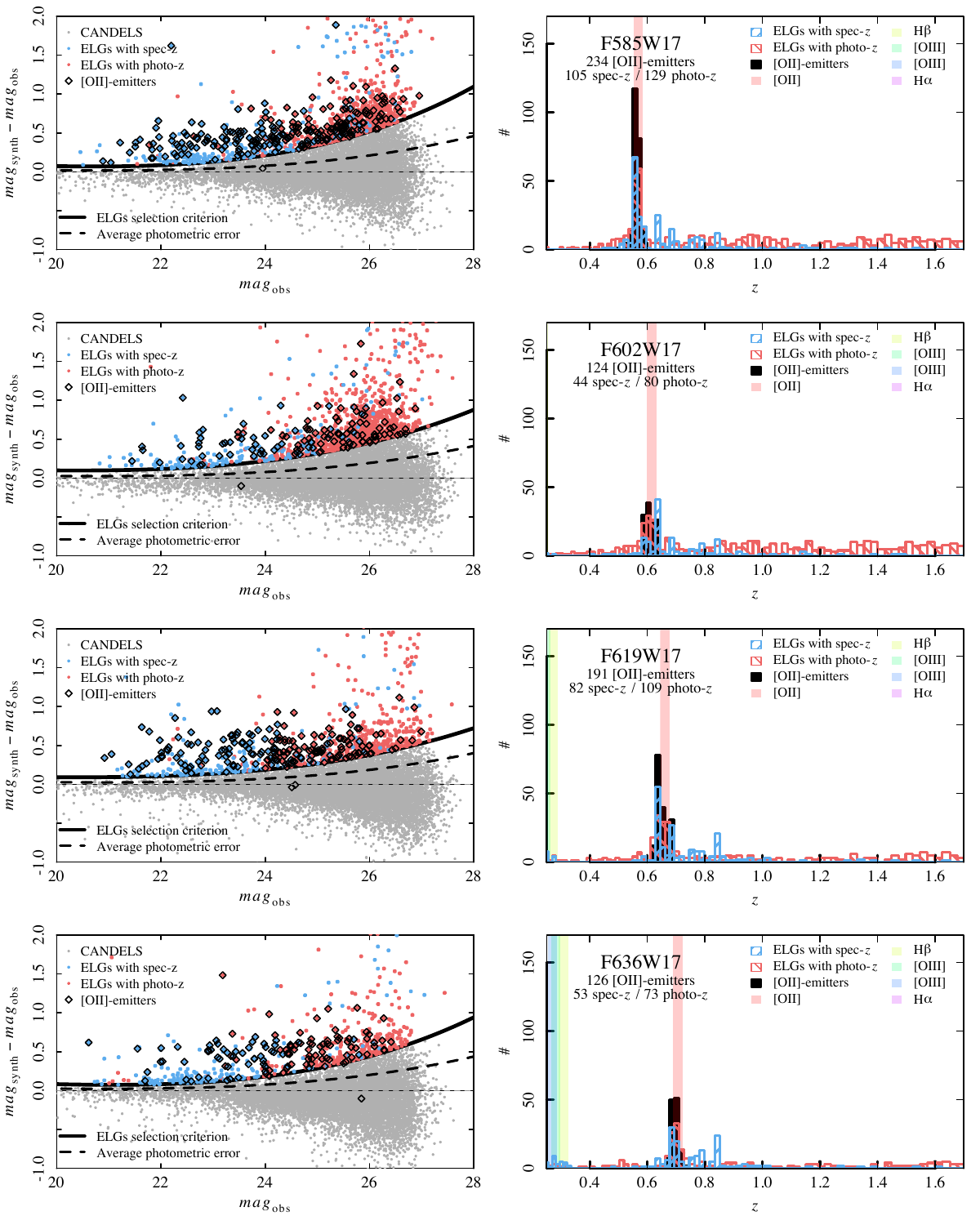}
\caption{Continued.}
\end{figure*}
\begin{figure*}
\addtocounter{figure}{-1}
\includegraphics[page=3,width=\linewidth]{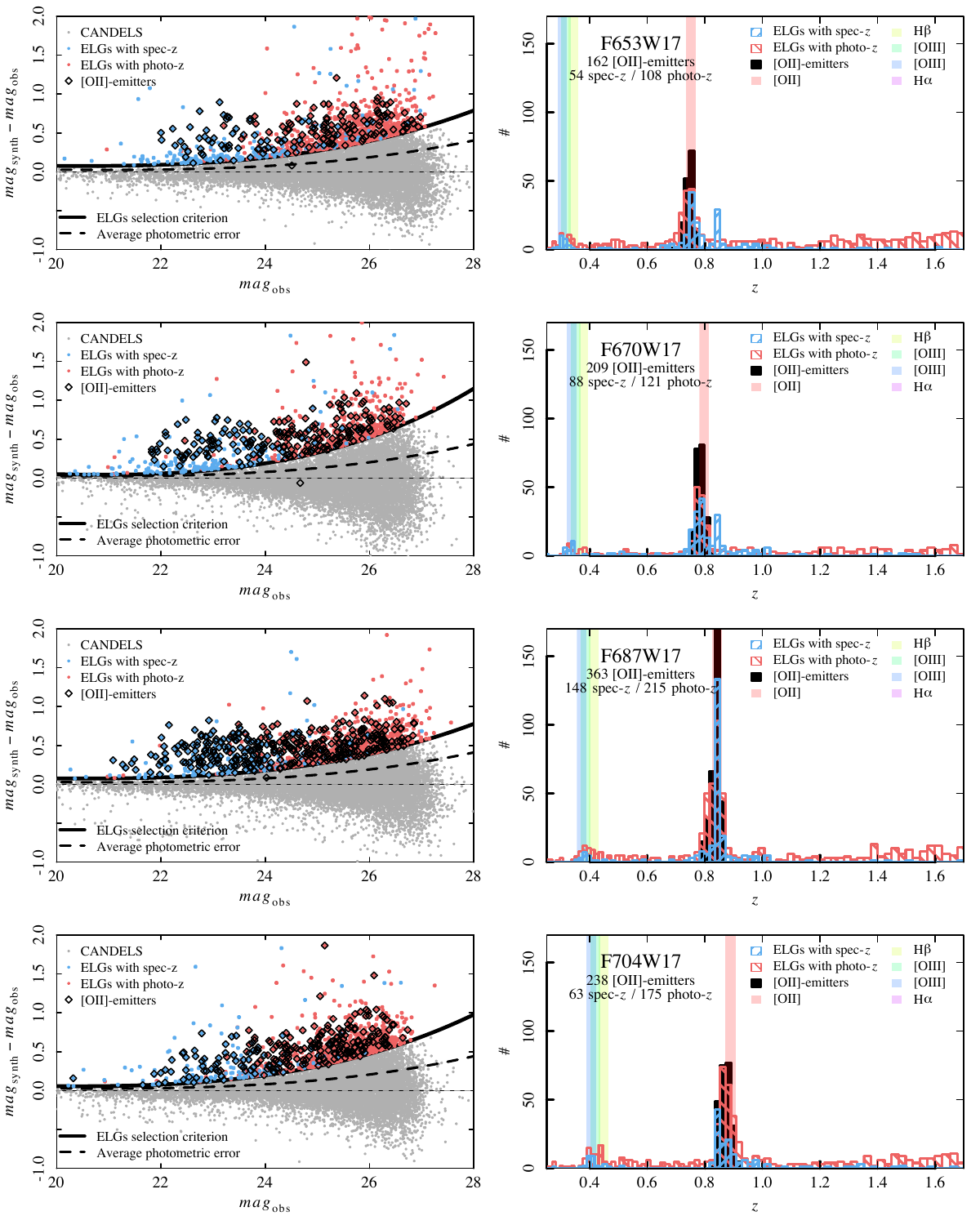}
\caption{Continued.}
\end{figure*}
\begin{figure*}
\addtocounter{figure}{-1}
\includegraphics[page=4,width=\linewidth]{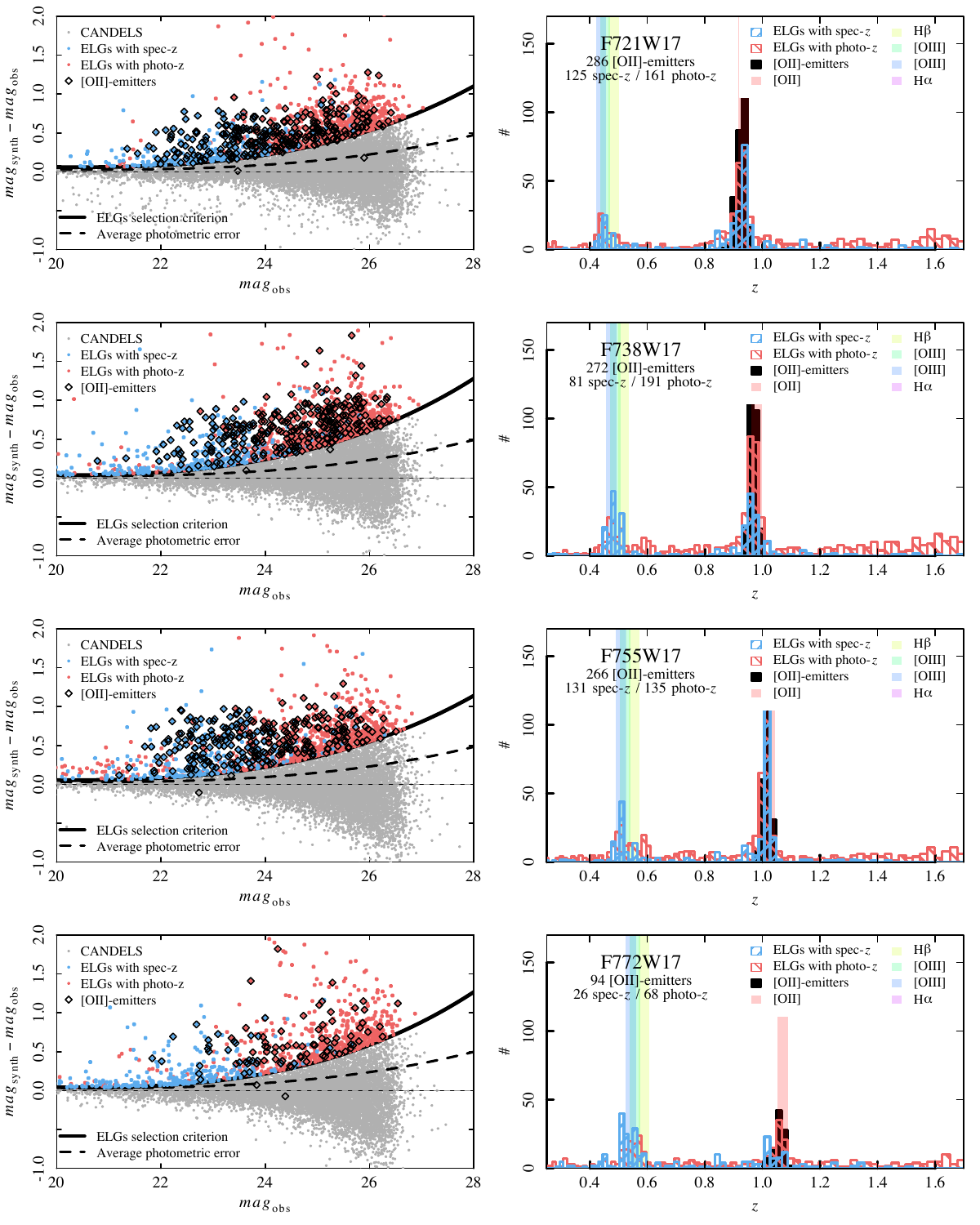}
\caption{Continued.}
\end{figure*}
\begin{figure*}
\addtocounter{figure}{-1}
\includegraphics[page=5,width=\linewidth]{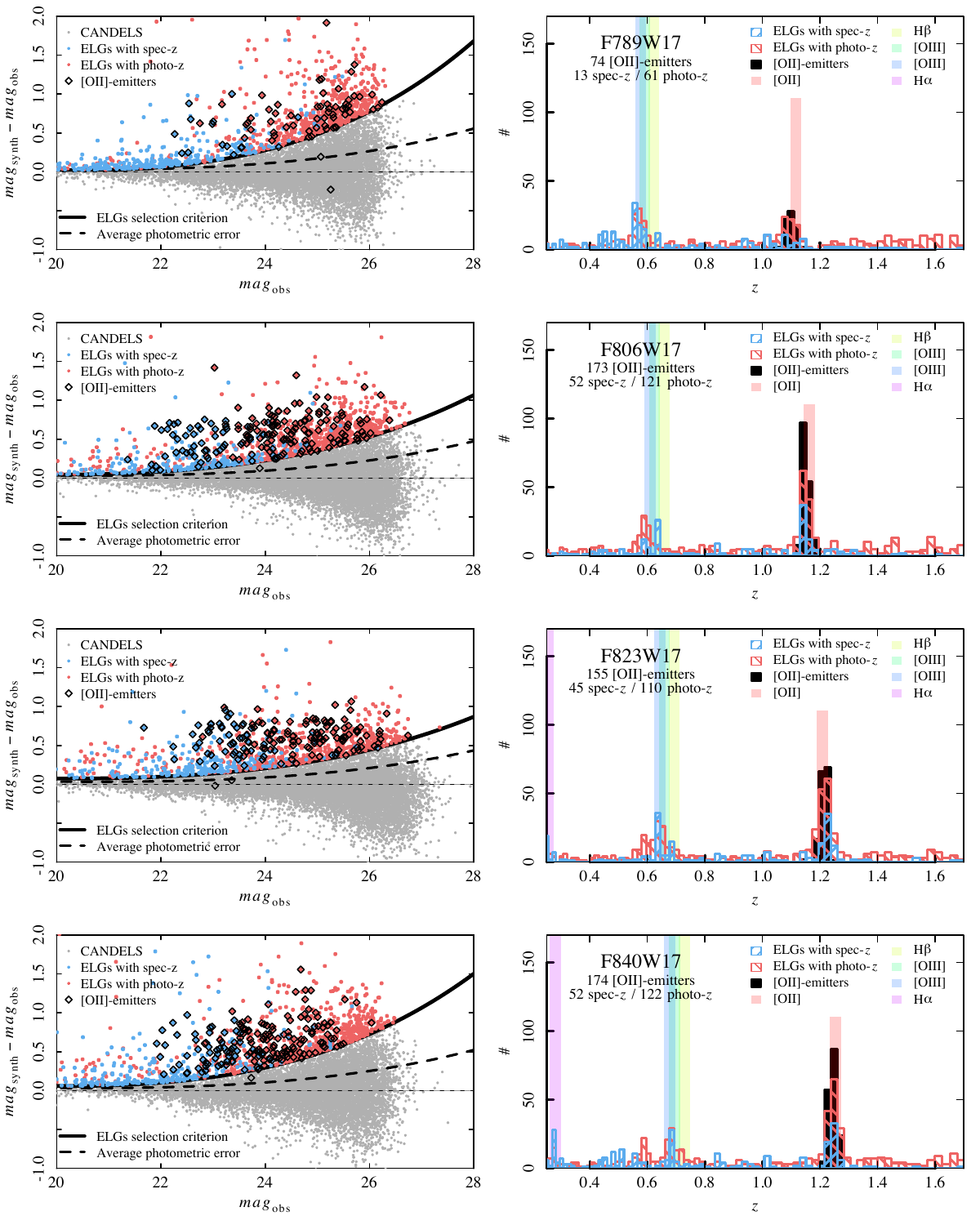}
\caption{Continued.}
\end{figure*}
\begin{figure*}
\addtocounter{figure}{-1}
\includegraphics[page=6,width=\linewidth]{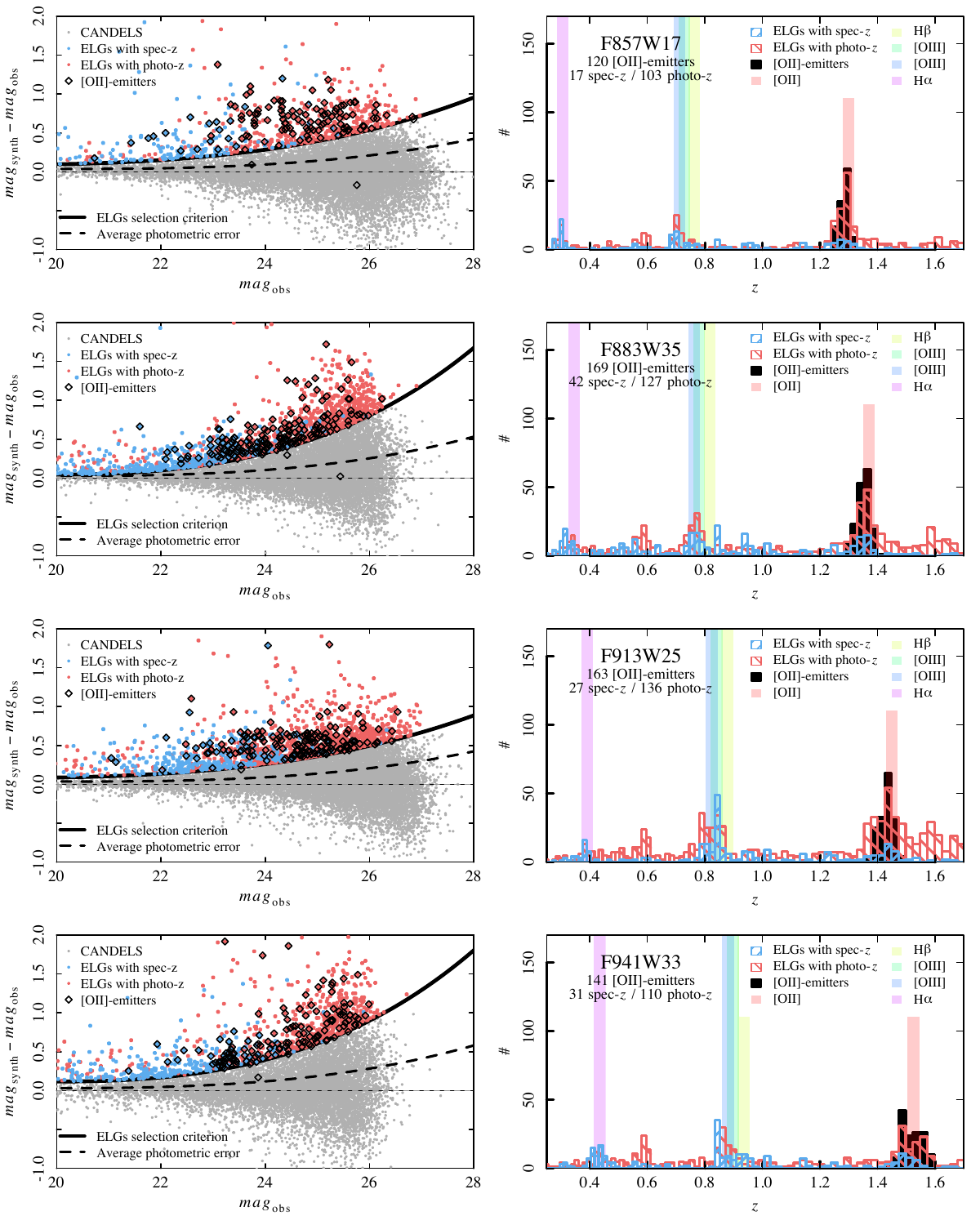}
\caption{Continued.}
\end{figure*}

\bsp	
\label{lastpage}
\end{document}